\documentclass[a4paper,fleqn,usenatbib]{mnras}

\usepackage{newtxtext,newtxmath}
\usepackage[T1]{fontenc}
\usepackage{ae,aecompl}

\usepackage{graphicx}	% Including figure files
\usepackage{amsmath}	% Advanced maths commands
\usepackage{amssymb}	% Extra maths symbols

\graphicspath{{.}{./images/}}
\newcommand\hco{HCO$^+$\:}
\newcommand\hcco{H$^{13}$CO$^+$\:}
\newcommand\hcoo{HC$^{18}$O$^+$\:}
\newcommand\dco{DCO$^+$\:}
\newcommand\dcco{D$^{13}$CO$^+$\:}

\title[3D modelling of HCO$^+$ in IRAS16293$-$2422]{3D modelling of HCO$^+$ and its isotopologues in the low-mass proto-star IRAS16293$-$2422}

\author[Qu\'enard et al.]{
D. Qu\'enard,$^{1,2}$\thanks{E-mail: d.quenard@qmul.ac.uk}
S. Bottinelli,$^{1}$
E. Caux$^{1}$
and V. Wakelam$^{3}$
\\
$^{1}$IRAP, Université de Toulouse, CNRS, UPS, CNES, Toulouse, France\\
$^{2}$School of Physics and Astronomy, Queen Mary University of London, Mile End Road, London E1 4NS, UK\\
$^{3}$Laboratoire d'astrophysique de Bordeaux, Univ. Bordeaux, CNRS, B18N, all\'ee Geoffroy Saint-Hilaire, 33615 Pessac, France
}

\date{Accepted XXX. Received YYY; in original form ZZZ}

\pubyear{2017}

\begin{document}
\label{firstpage}
\pagerange{\pageref{firstpage}--\pageref{lastpage}}
\maketitle

\begin{abstract}

Ions and electrons play an important role in various stages of the star formation process. By following the magnetic field of their environment and interacting with neutral species, they slow down the gravitational collapse of the proto-star envelope. This process (known as ambipolar diffusion) depends on the ionisation degree, which can be derived from the \hco abundance.
We present a study of \hco and its isotopologues (\hcco\!, \hcoo\!, \dco\!, and \dcco\!) in the low-mass proto-star IRAS16293$-$2422.
The structure of this object is complex, and the \hco emission arises from the contribution of a young NW-SE outflow, the proto-stellar envelope and the foreground cloud. We aim at constraining the physical parameters of these structures using all the observed transitions.
For the young NW-SE outflow, we derive $T_{\rm kin}=180-220$\,K and $n({\rm H_2})=(4-7)\times10^6$\,cm$^{-3}$ with an \hco abundance of $(3-5)\times10^{-9}$. Following previous studies, we demonstrate that the presence of a cold ($T_{\rm kin}$$\leqslant$30\,K) and low density ($n({\rm H_2})\leqslant1\times10^4$\,cm$^{-3}$) foreground cloud is also necessary to reproduce the observed line profiles. We have used the gas-grain chemical code \textsc{nautilus} to derive the \hco abundance profile across the envelope and the external regions where X(\hco\!)$\gtrsim1\times10^{-9}$ dominate the envelope emission. 
From this, we derive an ionisation degree of $10^{-8.9}\,\lesssim\,x(e)\,\lesssim\,10^{-7.9}$. The ambipolar diffusion timescale is $\sim$5 times the free-fall timescale, indicating that the magnetic field starts to support the source against gravitational collapse and the magnetic field strength is estimated to be $6-46\,\mu$G.

\end{abstract}


%%%%%%%%%%%%%%%%%%%%%%%%%%%%%%%%%%%%%%%%%%%%%%%%%%

%%%%%%%%%%%%%%%%% BODY OF PAPER %%%%%%%%%%%%%%%%%%

%________________________________________________________________

\section{Introduction}

The ionisation degree, defined as $x(e) = n(e)/n(\rm H_2)$ (with $n(e)$ the electron density and $n(\rm H_2)$ the H$_2$ density) plays an important role in regulating the star formation process \citep[e.g.][]{mouschovias1987, shu1987}. Indeed, through ambipolar diffusion, ions are separated from neutral species because of the ambient magnetic field. Due to collisions between ions and neutrals, the neutral matter, experiencing gravitational collapse, is slowed down from infalling into the central object. The timescale for ambipolar diffusion ($t_{\rm AD}$) depends on the ionisation degree ($t_{\rm AD}[yr] = 2.5\times10^{13}\,x(e)$, \citealp{spitzer1978, shu1987}). If the ratio of timescales between ambipolar diffusion and free-fall is close to one, the magnetic field is not playing any important role in preventing the collapse and it may lead to a situation where the object becomes gravitationally unstable. It is therefore important to determine the ionisation degree in order to understand the dynamics of a proto-stellar object.

\hco has been proven to be a good tracer of the degree of ionisation of a cloud because it indirectly probes the electron density when compared to other species \citep{caselli2002-1, caselli2002, caselli2002-2}. Moreover, around low-mass proto-stars, \hco is expected to be abundant owing to the relative simplicity of its formation, making it easy to detect and to study.

IRAS16293$-$2422 (hereafter IRAS16293) is a typical solar-type Class 0 low-mass proto-star located at 147.3\,pc \citep{ortiz-leon2017} embedded in the LDN1689N cloud within the $\rho$\,Ophiuchus complex. This source is well-studied due to its strong emission lines and its chemical richness \citep[e.g.][]{caux2011, jorgensen2011, jorgensen2016}. Many physical and chemical processes have been tested using observations toward this object, making it a ``template'' source in the past decades. The structure of this object is quite complex and several outflows have been detected and traced at multiple scales \citep{castets2001, stark2004, chandler2005, yeh2008, loinard2012, girart2014}. At small scales, it is composed of two distinct cores IRAS16293 A and IRAS16293 B separated by $\sim$5\arcsec~\citep{wootten1989, mundy1992}. Little is known on the physical properties of these cores although more and more studies aim at understanding their structure \citep[e.g.][]{chandler2005, rao2009, pineda2012, jacobsen2017} and chemical content \citep[e.g.][]{caux2011, jorgensen2011, jorgensen2016, persson2018}. To derive an accurate abundance of \hco (and therefore the ionisation degree), a complete description of the structure of the source is needed. Indeed, an estimation of the \hco abundance and some of its isotopologues in the envelope has already been obtained by \citet{van_dishoeck1995} and \citet{schoier2002} but without taking into account the complex structure of the source.

The goal of this study is twofold: we aim at giving better constraints on the 3D physical structure of IRAS16293 and to derive an accurate value of the \hco abundance in the different environments of this source. To do so, we use the observed emission of \hco and its isotopologues (\hcco\!, \hcoo\!, \dco\!, and \dcco\!), described in Sect. \ref{obs}. We then compare these observations to 3D radiative transfer modellings, presented in Sect. \ref{RT}. The 3D physical and chemical structure of the object is described in Sect. \ref{phys_chem} and our method in Sect. \ref{methods}. We derive constraints on the 3D physical structure of IRAS16293 in Sect. \ref{discussion} together with discussions on the parameters used in this study. From our modelling of the \hco abundance, we are finally able to determine the ionisation degree across the source. Concluding remarks are given in Sect. \ref{ccl}.

%________________________________________________________________

\section{Observations}\label{obs}

In this work, we are using data coming from two unbiased spectral surveys, \textit{i)} The IRAS16293 Millimetre And Sub-millimetre Spectral Survey, performed at the IRAM 30\,m ($80-265$\,GHz) and JCMT-15m ($330-370$\,GHz) telescopes between January 2004 and August 2006, and APEX-12m ($265-330$\,GHz) telescope between June 2011 and August 2012 (TIMASSS, \citealp{caux2011}) and \textit{ii)} the HIFI guaranteed time Key Program CHESS \citep{ceccarelli2010}. The HIFI data presented in this article are part of a survey observed between March 2010 and April 2011 providing full spectral coverage of bands 1a ($480-560$\,GHz; obsid 1342191499), 1b ($560-640$\,GHz; obsid 1342191559), 2a ($640-720$\,GHz; obsid 1342214468), 2b ($720-800$\,GHz; obsid 1342192332), 3a ($800-880$\,GHz; obsid 1342214308), 3b ($880-960$\,GHz; obsid 1342192330), 4a ($960-1040$\,GHz; obsid 1342191619), 4b ($1040-1120$\,GHz; obsid 1342191681), and 5a ($1120-1200$\,GHz; obsid 1342191683). The HIFI Spectral Scan Double Beam Switch (DBS) observing mode with optimisation of the continuum was used with the HIFI acousto-optic Wide Band Spectrometer (WBS), providing a spectral resolution of 1.1\,MHz ($\sim$0.6\,km\,s$^{-1}$ at 500\,GHz and $\sim$0.3\,km\,s$^{-1}$ at 1\,THz) over an instantaneous bandwidth of $4\times1$\,GHz \citep{roelfsema2012}. 

For the TIMASSS survey, the observed coordinates were $\alpha_{2000} = 16^{\rm{h}}32^{\rm{m}}22^{\rm{s}}\!.6$, $\delta_{2000}=-24^\circ28\arcmin33\arcsec$ while they were $\alpha_{2000} = 16^{\rm{h}}32^{\rm{m}}22^{\rm{s}}\!.75$, $\delta_{2000}=-24^\circ28\arcmin34.2\arcsec$ for the HIFI observations. The difference in the aimed positions has been carefully taken into account in this work. For both surveys, the DBS reference positions were situated $3\arcmin$ apart from the source. Table \ref{lines_obs} summarises the observation parameters.

The data processing of the TIMASSS survey has been extensively described in \citet{caux2011}. The HIFI data have been processed using the standard HIFI pipeline up to frequency and intensity calibrations (level 2) with the ESA-supported package HIPE 12 \citep{ott2010}. Using a standard routine developed within the HIFI ICC (Instrument Control Center), \textit{flagTool}, spurs not automatically detected by the pipeline have been tagged and removed. Then, the HIPE tasks \textit{fitHifiFringe} and \textit{fitBaseline} were used to remove standing waves and to fit a low-order polynomial baseline to line-free channels. Finally, sideband deconvolution was performed with the dedicated HIPE task \textit{doDeconvolution}.

The spectra observed in both horizontal and vertical polarisation were of similar quality, and averaged to lower the noise in the final spectrum, since polarisation is not a concern for the presented analysis. The continuum values obtained from running \textit{fitBaseline} are well fitted by polynomials of order 3 over the frequency range of the whole band. The single side band continuum derived from the polynomial fit at the considered frequencies (Table \ref{lines_obs}) was added back to the continuum-free spectra. Intensities were then converted from antenna to main-beam temperature scale using a forward efficiency of 0.96 and the (frequency-dependent) beam-efficiency shown in Table 1. Peak intensities are reported in Table \ref{lines_obs} together with the spectroscopic and observing parameters of the transitions used in this work.

\begin{table*}\caption{Parameters for the observed \hco\!, \hcco\!, \hcoo\!, \dco\!, and \dcco lines.}\label{lines_obs}
{\scriptsize
\begin{tabular}{ccrrcrccccccc}
\hline
\hline
Molecule&Transition&Freq.&$E_{\rm up}$&A$_{ij}$&rms&V$_{\textsc{lsr}}$\,$^{4}$&FWHM\,$^{4}$&$T_{\rm mb}$$^{5}$&$\int{T_{\rm mb}dv}$\,$^{5}$&Telescope&Beam&$\eta_{\rm mb}$\\
&$J_{\rm up}-J_{\rm low}$&(GHz)&(K)&(s$^{-1}$)&(mK)&(km\,s$^{-1}$)&(km\,s$^{-1}$)&(K)&(K\,km\,s$^{-1}$)&&size ($\arcsec$)&\\
\hline
HCO$^+$
&$1-0$\,$^{1}$&89.189&4&4.19$\times10^{-5}$&8.1&$-$&$-$&$-$&$15.3\pm1.69$&IRAM&27.8&0.81 \\
&$3-2$\,$^{1}$&267.558&26&1.45$\times10^{-3}$&55.8&$-$&$-$&$-$&$44.4\pm4.44$&APEX&23.5&0.75 \\
&$4-3$\,$^{1}$&356.734&43&3.57$\times10^{-3}$&26.1&$-$&$-$&$-$&$67.6\pm6.76$&JCMT&13.9&0.64 \\
&$6-5$&535.062&90&1.25$\times10^{-2}$&11.5&$3.52\pm0.03$&$2.43\pm0.08$&$7.73\pm0.80$&$21.1\pm2.11$&HIFI&39.7&0.62 \\
&$7-6$&624.208&120&2.01$\times10^{-2}$&12.4&$3.48\pm0.02$&$2.55\pm0.04$&$7.85\pm0.79$&$21.6\pm2.16$&HIFI&34.0&0.62 \\
&$8-7$&713.341&154&3.02$\times10^{-2}$&27.7&$3.55\pm0.02$&$2.86\pm0.04$&$6.50\pm0.65$&$20.1\pm2.01$&HIFI&29.7&0.65 \\
&$9-8$&802.458&193&4.33$\times10^{-2}$&37.5&$3.70\pm0.02$&$3.46\pm0.04$&$5.28\pm0.79$&$17.9\pm2.69$&HIFI&26.4&0.63 \\
&$10-9$&891.557&235&5.97$\times10^{-2}$&37.5&$3.83\pm0.01$&$3.88\pm0.03$&$4.42\pm0.66$&$15.6\pm2.34$&HIFI&23.8&0.63 \\
&$11-10$&980.636&282&7.98$\times10^{-2}$&44.6&$3.88\pm0.01$&$4.27\pm0.03$&$3.63\pm0.54$&$12.4\pm1.86$&HIFI&21.6&0.64 \\
&$12-11$&1069.694&334&1.04$\times10^{-1}$&63.3&$3.87\pm0.02$&$4.41\pm0.05$&$2.64\pm0.53$&$9.20\pm1.84$&HIFI&19.8&0.64 \\
&$13-12$&1158.727&389&1.33$\times10^{-1}$&147.2&$4.04\pm0.04$&$4.77\pm0.11$&$2.13\pm0.43$&$8.29\pm1.66$&HIFI&18.3&0.59 \\
\\
H$^{13}$CO$^{+}$&$1-0$&86.754&4&3.85$\times10^{-5}$&6.5&$4.17\pm0.01$&$2.13\pm0.21$&$1.91\pm0.21$&$4.13\pm0.45$&IRAM&28.5&0.81 \\
&$2-1$&173.507&12&3.70$\times10^{-4}$&22.1&$4.48\pm0.03$&$2.05\pm0.06$&$3.74\pm0.63$&$7.73\pm1.31$&IRAM&14.3&0.69 \\
&$3-2$&260.255&25&1.34$\times10^{-3}$&14.2&$3.61\pm0.02$&$3.04\pm0.04$&$3.19\pm0.54$&$9.98\pm1.70$&IRAM&9.50&0.53 \\
&$4-3$&346.998&42&3.29$\times10^{-3}$&22.1&$3.52\pm0.03$&$2.36\pm0.06$&$3.52\pm0.63$&$7.65\pm1.38$&JCMT&14.3&0.64 \\
&$6-5$&520.460&87&1.15$\times10^{-2}$&9.2&$4.02\pm0.02$&$3.21\pm0.05$&$0.35\pm0.04$&$1.02\pm0.10$&HIFI&40.8&0.62 \\
&$7-6$&607.175&117&1.85$\times10^{-2}$&11.2&$4.06\pm0.06$&$3.00\pm0.14$&$0.24\pm0.04$&$0.67\pm0.10$&HIFI&35.0&0.62 \\
&$8-7$&693.876&150&2.78$\times10^{-2}$&22.0&$4.18\pm0.11$&$3.31\pm0.27$&$0.14\pm0.04$&$0.29\pm0.09$&HIFI&30.6&0.65 \\
&$9-8$&780.563&187&3.99$\times10^{-2}$&27.2&$-$&$-$&$-$&$< 0.210$&HIFI&27.2&0.65 \\
\\
HC$^{18}$O$^{+}$&$1-0$&85.162&4&3.64$\times10^{-5}$&5.9&$4.17\pm0.05$&$1.96\pm0.13$&$0.18\pm0.03$&$0.35\pm0.04$&IRAM&29.1&0.81 \\
&$2-1$&170.323&12&3.50$\times10^{-4}$&19.6&$4.34\pm0.05$&$1.66\pm0.12$&$0.42\pm0.08$&$0.70\pm0.12$&IRAM&14.5&0.69 \\
&$3-2$\,$^{3}$&255.479&25&1.27$\times10^{-3}$&9.7&$3.57\pm0.27$&$2.43\pm0.71$&$0.11\pm0.06$&$0.28\pm0.15$&IRAM&9.7&0.54 \\
&$4-3$&340.631&41&3.11$\times10^{-3}$&16.2&$3.96\pm0.08$&$3.11\pm0.21$&$0.36\pm0.07$&$1.12\pm0.20$&JCMT&14.5&0.64 \\
&$6-5$&510.910&86&1.09$\times10^{-2}$&10.6&$4.27\pm0.11$&$2.31\pm0.27$&$0.06\pm0.04$&$0.12\pm0.06$&HIFI&41.6&0.62 \\
&$7-6$&596.034&114&1.75$\times10^{-2}$&13.7&$-$&$-$&$-$&$< 0.106$&HIFI&35.8&0.62 \\
\\
DCO$^+$&$2-1$&144.077&10&2.12$\times10^{-4}$&13.7&$4.32\pm0.01$&$1.42\pm0.01$&$3.57\pm0.61$&$5.45\pm0.93$&IRAM&17.2&0.73\\
&$3-2$&216.113&21&7.66$\times10^{-4}$&17.8&$4.41\pm0.03$&$2.91\pm0.09$&$1.74\pm0.30$&$5.67\pm0.97$&IRAM&11.5&0.62 \\
&$4-3$&288.144&35&1.88$\times10^{-3}$&8.90&$4.20\pm0.03$&$2.24\pm0.06$&$1.24\pm0.25$&$2.87\pm0.57$&APEX&21.8&0.75 \\
&$5-4$&360.170&52&3.76$\times10^{-3}$&20.9&$3.86\pm0.04$&$2.51\pm0.10$&$1.16\pm0.21$&$2.57\pm0.46$&JCMT&13.7&0.64 \\
&$7-6$&504.200&97&1.06$\times10^{-2}$&10.9&$4.02\pm0.09$&$3.03\pm0.22$&$0.08\pm0.02$&$0.22\pm0.05$&HIFI&42.1&0.62 \\
&$8-7$&576.202&124&1.59$\times10^{-2}$&9.1&$-$&$-$&$-$&< 0.396\,$^{2}$&HIFI&36.8&0.62 \\
\\
D$^{13}$CO$^+$&$2-1$&141.465&10&2.00$\times10^{-4}$&11.7&$4.38\pm0.07$&$1.27\pm0.16$&$0.21\pm0.06$&$0.27\pm0.05$&IRAM&17.5&0.73 \\
&$3-2$&212.194&20&7.25$\times10^{-4}$&6.6&$4.43\pm0.26$&$2.69\pm0.61$&$0.08\pm0.03$&$0.22\pm0.04$&IRAM&11.7&0.63 \\
&$4-3$&282.920&34&1.78$\times10^{-3}$&7.10&$4.55\pm0.14$&$3.05\pm0.34$&$0.07\pm0.02$&$0.21\pm0.04$&APEX&22.2&0.75 \\
&$5-4$&353.640&51&3.56$\times10^{-3}$&27.3&$-$&$-$&$-$&< 0.211&JCMT&14.0&0.64 \\
\hline
\multicolumn{13}{l}{\textbf{Notes.} (1) Lines showing a strong self-absorption profile.}\\
\multicolumn{13}{l}{(2) Transition blended with CO (J = 5 $\rightarrow$ 4).}\\
\multicolumn{13}{l}{(3) Transition badly calibrated (see \citealp{caux2011}).}\\
\multicolumn{13}{l}{(4) The error bars only correspond to a statistical error estimated from a Gaussian fit.}\\
\multicolumn{13}{l}{(5) The error bars are calculated following: $\sigma_{\rm tot}=\sqrt{\sigma_{\rm cal}^2 + \sigma_{\rm stat}^2}$ with $\sigma_{\rm cal}$ the calibration error taken from \citet{caux2011} and}\\
\multicolumn{13}{l}{\citet{ceccarelli2010} and $\sigma_{\rm stat}$ the statistical error estimated from a Gaussian fit.}
\end{tabular}
}
\end{table*}

%________________________________________________________________

\section{Radiative transfer modelling}\label{RT}

We have first tried to fit the detected lines using simpler radiative transfer models: Boltzmann diagrams, LTE modelling, and LVG calculations. None of these methods gave satisfactory results, thus, in order to derive the line profile of the studied molecular transitions and the continuum emission, we have used \textsc{lime}, a 3D non-LTE radiative transfer code \citep{brinch2010}. 
To describe the input 3D physical model of IRAS16293 and set the different parameters of \textsc{lime} we have used \textsc{gass} (Generator of Astrophysical Sources Structures, \citealp{quenard2017}). \textsc{gass} is a user friendly interface that allows to create, manipulate, and mix one or several different physical structures such as spherical sources, discs, and outflows. \textsc{gass} is fully adapted to \textsc{lime} and it produces output models than can be directly read by \textsc{lime}. A complete description of the procedure that creates the different structures of the physical model is given in \citet{quenard2017}.

Once output data cubes have been generated by \textsc{lime} they have been post-processed using \textsc{gass}. For single-dish observations, the treatment consists in convolving the cube with the beam size of the desired telescope and to plot the predicted spectrum in main beam temperature as a function of the velocity for each observed frequency. The cube is built with a better spectral resolution (set to 100\,m\,s$^{-1}$ for all models) than the observations but the predicted spectra are resampled at the same spectral resolution as that of the observations. We carefully take into account the different telescopes source pointings in the convolution by using the appropriate positions given in Sect. \ref{obs}.

All \hco and its isotopologues (except \dcco\!) collision files have been taken from the Leiden Atomic and Molecular Database\footnote{\url{http://home.strw.leidenuniv.nl/~moldata/}} (LAMDA, \citealp{schoier2005}). For each molecule, we have updated the spectroscopic values implemented in these collision files with the newest spectroscopic data taken from the Cologne Database for Molecular Spectroscopy\footnote{\url{http://www.astro.uni-koeln.de/cdms/}} (CDMS, \citealp{muller2005}). The collisional rates are taken from \citet{flower1999} and were calculated for temperatures in the range from 10 to 400\,K including energy levels up to $\rm{J}=20$ for collisions with H$_2$. Since the \dcco file does not exist, we have created it from the CDMS database by considering that the collisional rates are the same as for \dco\!.

%________________________________________________________________

\section{Physical and chemical structure}\label{phys_chem}

\begin{figure*}
	\centering
	\includegraphics[width=\hsize, clip=true, trim= 0 0 0 0]{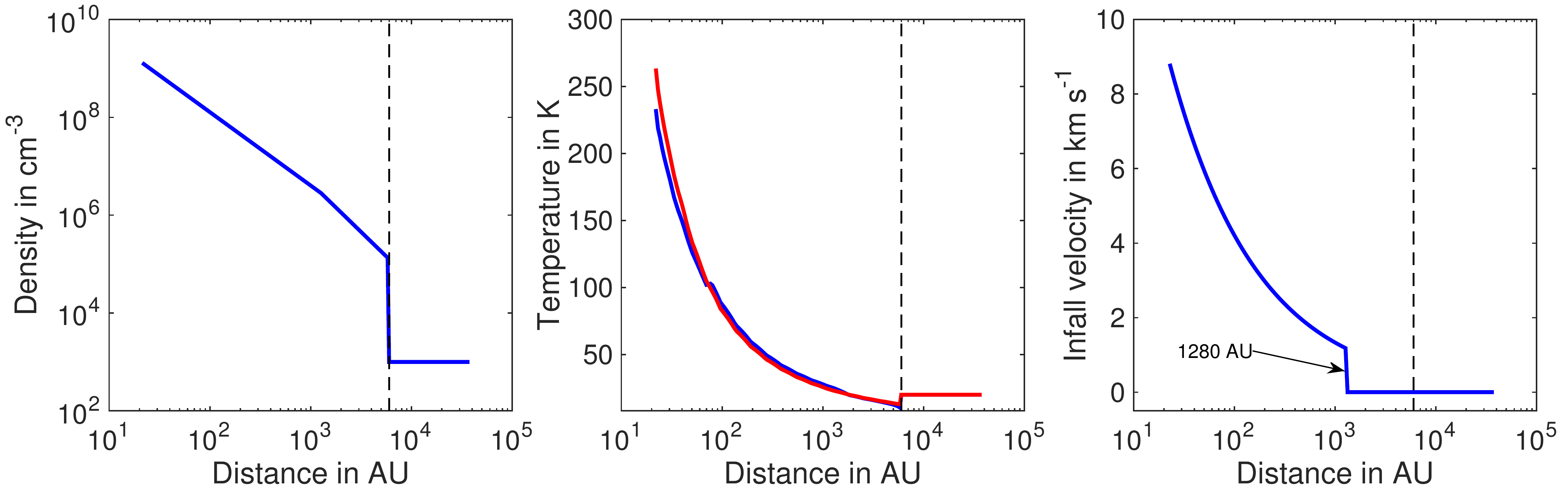}
	\caption{\textit{Left panel:} Density profile of IRAS16293 used in this study as a function of the radius. \textit{Middle panel:} Gas (blue) and dust (red) temperature profiles of IRAS16293 used in this study as a function of the radius. \textit{Right panel:} Absolute value of the radial infall velocity as a function of the radius. In all panels, the black dotted line shows the $\rm{R}=6000$\,au limit of the foreground cloud.}
	\label{struct_profile}
\end{figure*}

The physical structure of IRAS16293 is very complex with multiple outflows, multiple sources, and an envelope. To reproduce the observed \hco emission and line profiles we have modelled in 3D the different structures that contribute to the overall emission. For that, we define the physical structure of each component and their respective \hco abundance profile, as described in the following sub-sections.

%________________________________________________________________

\vspace{-0.2cm}

\subsection{The envelope model}\label{chem_env}

\subsubsection{Physical profile}

\citet{crimier2010} have derived the physical structure of the source (H$_2$ density, gas and dust temperature profiles of the envelope).
This physical profile has been used in several physical and chemical studies of the source \citep[e.g.][]{hily-blant2010, vastel2010, coutens2012, bottinelli2014, jaber2014, wakelam2014, lopez-sepulcre2015, majumdar2016} and we base our physical structure on the same definition. The Shu-like density distribution described by \citet{crimier2010} is:
\begin{eqnarray}
	n(r) &=& n(r_{\rm in})\times\left(\frac{r_{\rm in}}{r}\right)^{1.5} ~ \textnormal{if} ~ r < r_{\rm inf},\\
	n(r) &=& n(r_{\rm in})\times\left(\frac{r_{\rm in}}{r}\right)^2 ~ \textnormal{if} ~ r \geq r_{\rm inf},
\end{eqnarray}
where $n(r_{\rm in})$ is the density at $r_{\rm in}$, the inner radius of the envelope, and $r_{\rm inf}$ refers to the radius where the envelope begins to collapse, marking a change in the slope of the density profile \citep{shu1977}. From their best result, \citet{crimier2010} derived $r_{\rm inf}=1280$\,au, $r_{\rm in}=22$\,au and $n(r_{\rm in})=1.23\times10^9$\,cm$^{-3}$.
Using these parameters and equations (1) and (2), we performed a spline interpolation of the profiles down to 1\,au, for the sake of the radiative transfer modelling. The density profiles, gas and dust temperature profiles are shown in the left and middle panels of Fig. \ref{struct_profile}, respectively.
One must note that the presence of multiple sources and proto-planetary discs in the core of the envelope (hot corinos) cannot be taken into account in the \textsc{DUSTY} model of \citet{crimier2010}. Thus, the inner structure of the source (< 600 au) is still open to discussions. However, the abundance of \hco in this region of the envelope does not contribute significantly to the total emission of the line, as discussed in Sect. \ref{res_chem_env}.

The velocity field of the source is based on the standard infall law in which the envelope is free-falling inside $r_{\rm inf}$, and considered to be static outside (thus the infall velocity is set to 0). Hence, the velocity field is described by the following equations:
\begin{eqnarray}
	V_{\rm{inf}}(r) &=& V_{\rm{inf}}(r_{\rm in}) ~ \textnormal{if} ~ r \leq r_{\rm in},\\
	V_{\rm{inf}}(r) &=& \sqrt{\frac{2\,G\,M_{\star}}{r}} ~ \textnormal{if} ~ r_{\rm in} < r < r_{\rm inf},\\
	V_{\rm{inf}}(r) &=& 0 ~ \textnormal{if} ~ r \geq r_{\rm inf},
\end{eqnarray}
where $V_{\rm{inf}}$ is the infall velocity, $G$ the gravitational constant, $M_{\star}$ the mass of the central object, and $r$ the distance from the central object (see right panel of Fig. \ref{struct_profile}). This formalism of the velocity field as already been used in a previous study of the source \citep{coutens2012}. For IRAS16293, the mass of the source A dominates the system and we set $M_{\star}=1\,M_\odot$ \citep{ceccarelli2000, ceccarelli2000-1, schoier2002, pech2010, coutens2012}. We have checked that a variation of the mass between 0.8 and 1.5 solar masses only changes line widths by up to 10\%. For smaller or larger value, the modelled \hco lines are too narrow or too broad, respectively. The envelope is supposed to be centred on IRAS16293 A since it is the more massive component of the binary system.

\subsubsection{Chemical modelling of \hco}\label{chem_hco}

We have used the gas-grain chemical code \textsc{nautilus} \citep[e.g.][]{ruaud2016} to estimate the radial abundance profile of \hco in the envelope of IRAS16293. \textsc{nautilus} computes the evolution of the species' abundances as a function of time in the gas-phase and on grain surfaces. A large number of gas-phase processes are included in the code: bimolecular reactions (between neutral species, between charged species and between neutral and charged species) and unimolecular reactions, i.e. photo-reactions with direct UV photons and UV photons produced by the de-excitation of H$_2$ excited by cosmic-ray particles (Pratap \& Tarafdar mechanism), photo-desorption, and direct ionisation and dissociation by cosmic-ray particles. The interactions of the gas phase species with the interstellar grains are: sticking of neutral gas-phase species to the grain surfaces, evaporation of the species from the surfaces due to the temperature, the cosmic-ray heating and the exothermicity of the reactions at the surface of the grains (a.k.a chemical desorption). The species can diffuse and undergo reactions using the rate equation approximation at the surface of the grains \citep{hasegawa1992}. Details on the processes included in the model can be found in \citet{ruaud2016}. Note that we have used \textsc{nautilus} in its two-phase model, meaning that there is no distinction between the surface and the bulk of the mantle of the grains.
The gas-phase reactions are based on the \texttt{kida2015.uva.2014} network\footnote{\url{http://wakelam2015.obs.u-bordeaux1.fr/networks.html}} \citep[see][]{wakelam2015} while the surface reactions are based on the \citet{garrod2006} network. The full network contains 736 species (488 in the gas-phase and 248 at the surface of the grains) and 10466 reactions (7552 pure gas-phase reactions and 2914 reactions of interactions with grains and reactions at the surface of the grains).
For this study we adopted the initial atomic abundances (with respect to the total proton density $n_{\rm H}$) given in \citet{hincelin2011} with an additional atomic abundance of 6.68$\times10^{-9}$ for fluorine \citep{neufeld2005}. The carbon and oxygen abundances are respectively 1.7$\times10^{-4}$ and 3.3$\times10^{-4}$ leading to a C/O ratio of $\sim$0.5.

The chemical modelling of the envelope is done in two steps, as explained in \citet{quenard2017-1}. Initially, it starts by considering a static 0D parental cloud extended up to $r=4\times10^4$\,au with an initial gas temperature varying from 10 to 30\,K (see Sect. \ref{methods}) and a high visual extinction to prevent any photo-dissociation to occur.
The second step starts with the final abundances of the parental cloud step. In this phase, we consider the 1D physical structure of the envelope, supposed static as a function of time (see Sect. \ref{res_chem_env}). Abundances of species as a function of the radius are output for different ages of the proto-star, up to $1\times10^5$\,yr. The cosmic ray ionisation rate $\zeta$ is supposed to be the same as the one used for the parental cloud step. The visual extinction in the envelope is a function of the atomic hydrogen column density $\rm{N_H}$, calculated from the H$_2$ density profile:
\begin{equation}\label{eq_av}
	A_V=\frac{\rm{N_H}}{1.59\times10^{21}\,\rm{cm}^{-2}}.
\end{equation}
We also take into account the additional extinction from the foreground cloud in which we assume that the envelope is embedded (see \S1 of Sect. \ref{phy_chem_PDR}).

The chemical reaction network of \hco depends strongly on H$_3^+$ since its primary formation route is:
\begin{eqnarray}\label{hco_form}
	\mathrm{H_3^+ + CO} &\longrightarrow& \mathrm{HCO^+ + H_2.}
\end{eqnarray}

H$_3^+$ is formed from H$_2$ and strongly depends on the cosmic ray (CR) ionisation rate $\zeta$:
\begin{eqnarray}
	\mathrm{H_2 + CR} &\longrightarrow& \mathrm{H_2^+ + e^-,}\\
	\mathrm{H_2^+ + H_2} &\longrightarrow& \mathrm{H_3^+ + H.}
\end{eqnarray}

In Sect. \ref{res_chem_env} we show the significance of the cosmic ray ionisation rate in the abundance profiles of \hco in the envelope, and hence the contribution of this structure to the total emission of this species.

%________________________________________________________________

\subsection{The foreground cloud}\label{pdr_chem}

IRAS16293 is embedded in the remnants of its parental cloud, forming a foreground layer in the line of sight. This cloud has been studied by \citet{coutens2012} and \citet{wakelam2014} to analyse the deuteration in the source, and \citet{bottinelli2014} to investigate CH in absorption. Based on these studies, this cloud must be cold ($T_{\rm kin}$\,$\sim$\,10--30\,K) and not very dense ($n({\rm H_2})$$\sim$10$^3$--10$^5$\,cm$^{-3}$) with an $A_V$ of 1--4, similar to the physical conditions found in diffuse or translucent clouds \citep{hartquist1998}. The expected range of the \hco abundance is a few $10^{-10}-10^{-8}$ [unless stated differently, all abundances in this study are defined with respect to $n({\rm H}_2)$], depending on the temperature and on the degree of ionisation of the cloud \citep{lucas1996, hartquist1998, savage2004}. The V$_{\rm{LSR}}$ of the foreground cloud is supposed to be $4.2$\,km\,s$^{-1}$ (compared to the V$_{\rm{LSR}}=3.8$\,km\,s$^{-1}$ of IRAS16293), as derived by previous studies \citep{vastel2010, coutens2012, bottinelli2014}.

%________________________________________________________________

\subsection{The outflow model}\label{outflow_model}

The observed line shapes and intensities cannot be explained only with the contribution of the envelope of the source, particularly for high upper energy level transitions (e.g. $J_{{\rm up}}$>9). This effect has been also observed by \citet{gregersen1997} who performed a survey of \hco toward 23 Class 0 proto-stars. They determined from their \hco spectra (using \hco(4--3) and (3--2) transitions) that emission coming from bipolar outflows are contaminating the wings of the line, that might be confused with emission coming from the infalling envelope. Moreover, \hco is a molecule known to trace low-velocity entrained gas of outflows \citep[e.g.][]{sanchez-monge2013}.

A young NW-SE outflow ($\sim$400 yr) has been traced with SiO and CO emission \citep{rao2009, girart2014} using the SMA interferometer. \citet{rawlings2000} and \citet{rollins2014} have shown that young outflows can lead to an enhancement of the \hco abundance in a short period of time. Briefly, the interaction between the jet and/or the outflowing material and the surrounding quiescent (or infalling) gas is eroding the icy mantle of dust grains, desorbing the molecular materials in the gas phase (e.g. H$_2$O, CO, H$_2$CO, CH$_3$OH). Thanks to the photo-chemical processing induced by the shock-generated radiation field, this sudden enrichment of the gas-phase molecular abundances leads to the formation of many other molecules, such as \hco\!. \hco will be then destroyed by dissociative recombination or by interaction with water. Thus, we do not expect a high \hco abundance in old outflows but rather in young ones (<\,few hundred years old, \citealp{rawlings2000}) such as the NW-SE outflow detected in IRAS16293. \citet{rao2009} observed \hcco arising from the same region as this NW-SE outflow but they associated it to rotating material around IRAS16293 A rather than the outflow. Since the direction of this rotating material is roughly the same as the NW-SE outflow, it is more probable that the \hcco emission they observed is due to the recent enhancement of abundance. This conclusion is also supported by more recent \hcco SMA maps presented by \citet{jorgensen2011} (see their Fig. 19). Therefore, we have included this outflow in our 3D model together with the envelope.\\

We have considered an hourglass-like geometry for the outflow, as used by \citet{rawlings2004} in their study of \hco\!. This model is based on the mathematical definition given in \citet{visser2012}  implemented in \textsc{gass} (see \citealp{quenard2017} for more details on the outflow modelling). \citet{rao2009} and \citet{girart2014}, using SMA interferometric observations, derived the maximum extent of this outflow ($8\arcsec$), its inclination ($44^\circ$), dynamical age ($\sim400$\,yr), position angle ($145^\circ$), and velocity (V$_{{\rm outflow}}=15$\,km\,s$^{-1}$). Density and temperature are not really well constrained but, based on their SiO\,(8--7) emission, \citet{rao2009} suggested that this outflow is dense ($n({\rm H_2})\sim1\times10^7$\,cm$^{-3}$) and hot ($T_{{\rm kin}}\sim400$\,K). A correlation between SiO and \hco emission in outflows has already been observationally reported in high-mass star-forming regions \citep{sanchez-monge2013} hence we decided to consider that the \hco emission arises from a region with similar physical properties.

We aim at giving better constraints on the density and temperature of the outflow using all the \hco observations, thus we choose to only vary the gas temperature, the H$_2$ density, as well as the \hco abundance, all three considered to be constant as a function of the radius as a first approximation.

\begin{figure}
	\centering
	\includegraphics[width=\hsize, clip=true, trim= 0 0 0 0]{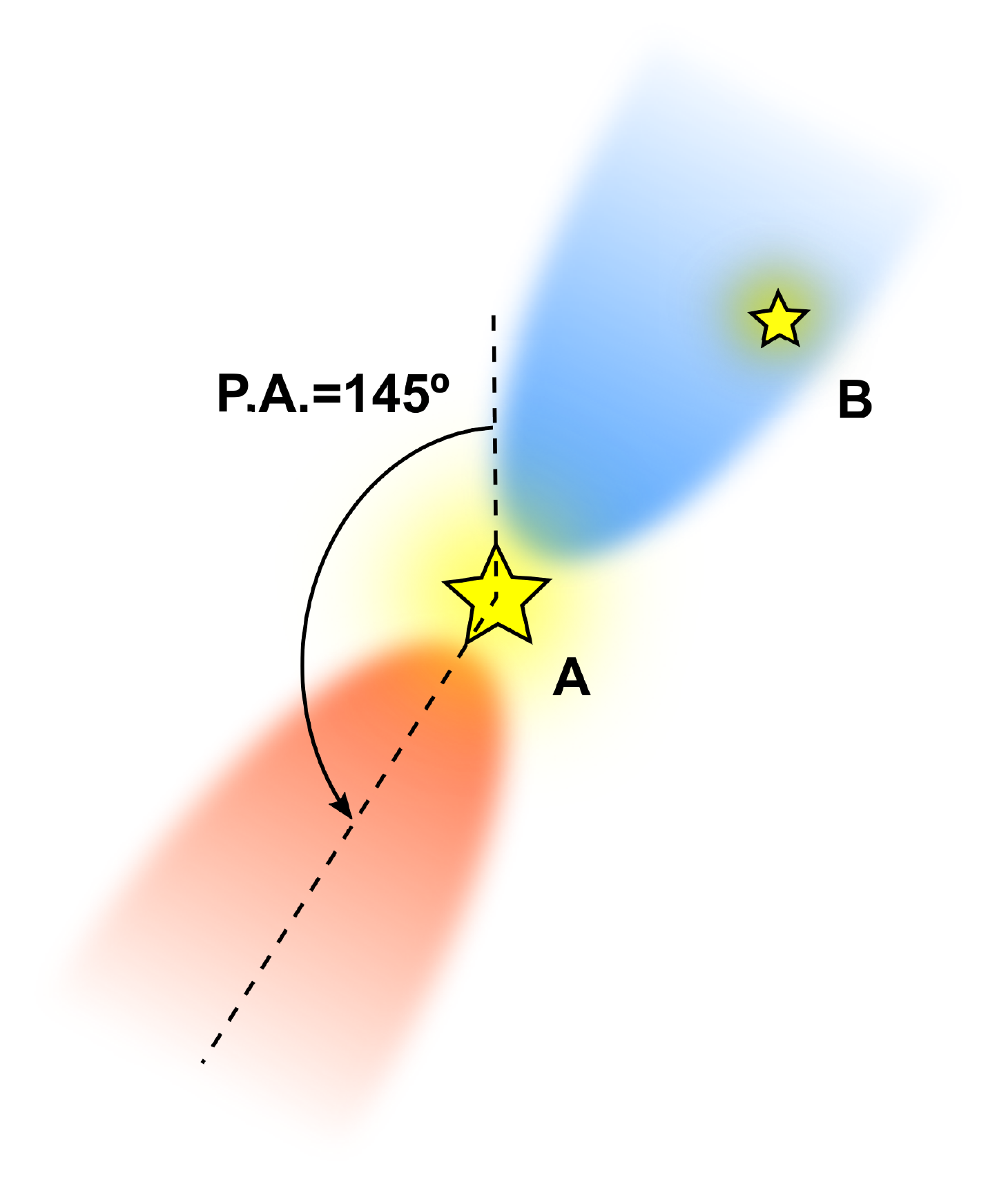}
	\caption{Sketch of the blueshifted (in blue) and redshifted (in red) NW-SE outflow emission. Both sources A and B (separated by 5\arcsec) are drawn but only source A is considered in the modelling.}
	\label{sketch}
\end{figure}

This outflow is quite young, collimated, and its low velocity suggests that the surrounding envelope is being pushed by the outflowing material. This kind of outflow-envelope interaction has already been observed and studied by \citet{arce2005, arce2006} for similar objects. Such interaction between the outflow and the envelope implies that there is no outflow cavity, as suggested by the interferometric observations, so we did not set it in the models.
Fig. \ref{sketch} presents a sketch of the outflow orientation and position in the model with respect to sources A and B.

%________________________________________________________________

\section{Methods}\label{methods}

One of the strengths of this study resides in the use of the data from the unbiased spectral surveys TIMASSS and CHESS (Sect. \ref{obs}), which provides us with a large number of transitions, spanning a wide range of upper energy levels ($4-389$\,K). Since the different structures contained in the physical model of the source span different temperature (and density) conditions, each structure is probed by a different set of transitions. For instance, the emission of the \hco low $J=1\rightarrow0$ transition (E$_{{\rm up}}=4$\,K) is more sensitive to the cold foreground cloud conditions while the $J_{{\rm up}}\gtrsim8$ transitions (E$_{{\rm up}}\gtrsim154$\,K) of the same molecule will preferentially help to constrain the outflow physical parameters, but poorly the envelope since the latter is much colder and is therefore contributing less to the total emission of these lines.

All 31 detected transitions of \hco and its isotopologues have been modelled using \textsc{gass} and \textsc{lime}. A grid of more than 5000 models have been calculated to constrain the physical properties of the envelope, foreground cloud and outflow. The range of tested values is shown in Table \ref{table_var}. To derive the best fit model, we have compared the predicted line fluxes of all lines with the observed values. Models where all line fluxes are falling within an error bars of 20\% with respect to the observed value are also shown in Table \ref{table_var}. These 20\% percent error models are here to show how restrictive the constraints found in our study are (or not) with respect to the best model.

\begin{table*}
	{
	\caption{Range of physical properties varied in this study}\label{table_var}
	\centering
	\begin{tabular}{cccc}
		\hline\hline
		Physical properties & Tested range & Best fit value & 20\% error bar\\
		\hline
		Age of the parental cloud & $10^5-10^6$\,yr & $1\times10^5$\,yr & $\leqslant3\times10^5$\,yr\\
		$n({\rm H_2})_{{\rm parental\,cloud}}$ & $1\times10^4-3\times10^{5}$\,cm$^{-3}$ & $3\times10^{4}$\,cm$^{-3}$ & $\leqslant1\times10^{5}$\,cm$^{-3}$\\
		$T_{{\rm parental\,cloud}}$ & $5-15$\,K & 10\,K & None\\
		$\zeta_{{\rm envelope}}$ & $1\times10^{-17}-1\times10^{-16}$\,s$^{-1}$ & $1\times10^{-16}$\,s$^{-1}$ & $\geqslant5\times10^{-17}$\,s$^{-1}$\\
		Age of the proto-star & up to $10^5$\,yr & $4\times10^4$\,yr & $\geqslant2\times10^4$\,yr\\
		\\
		$n({\rm H_2})_{{\rm foreground}}$ & $5\times10^2-3\times10^5$\,cm$^{-3}$ & $2\times10^3$\,cm$^{-3}$ & $5\times10^2-1\times10^4$\,cm$^{-3}$\\
   		$T_{{\rm kin, foreground}}$ & $10-30$\,K & 20\,K & None\\
		X(HCO$^+$)$_{{\rm foreground}}$ & $5\times10^{-11}-1\times10^{-6}$ & $1\times10^{-8}$ & $1\times10^{-9}-1\times10^{-7}$\\		
		\\
		$n({\rm H_2})_{{\rm outflow}}$ & $1\times10^6-1\times10^8$\,cm$^{-3}$ & $5.5\times10^6$\,cm$^{-3}$ & $(4-7)\times10^6$\,cm$^{-3}$\\
   		$T_{{\rm kin, outflow}}$ & $100-500$\,K & 200\,K & $180-220$\,K\\
		X(HCO$^+$)$_{{\rm outflow}}$ & $1\times10^{-10}-1\times10^{-7}$ & $4\times10^{-9}$ & $(3-5)\times10^{-9}$\\
		\hline
		\multicolumn{4}{l}{\textbf{Notes.} \textit{Top panel:} Parameters of the envelope. \textit{Middle panel:} Parameters of the foreground cloud.}\\
		\multicolumn{4}{l}{\textit{Bottom panel:} Parameters of the outflow.}
	\end{tabular}
	}
\end{table*}

In the following section we discuss the impact of the input parameters on the line profiles. As said above, each structure (envelope, outflow, foreground cloud) is probed by different \hco transitions, driven by the upper energy level of the transition. Therefore, to simplify the discussion on the impact of the input parameters, we have selected a subset of lines that best represent (visually) the variations of the parameters of each structure. For the envelope, the \hco(8--7) transition is the most sensitive to the variation of the envelope parameters while for the foreground cloud the low$-J$ \hco(1--0) transition is used and for the outflow the high$-J$ \hco(10--9) is selected.

%________________________________________________________________

\section{Results and discussions}\label{discussion}

The best model parameters we derived following our method are summarised in Table \ref{table_var}. A comparison between the observed line profiles and the best fit model for all the studied transitions of \hco\! is presented in Figure \ref{best_fit}.

%________________________________________________________________

\subsection{Chemistry of \hco in the envelope}\label{res_chem_env}

\begin{figure*}
	\centering
	\includegraphics[width=0.95\hsize, clip=true, trim= 30 72 0 0]{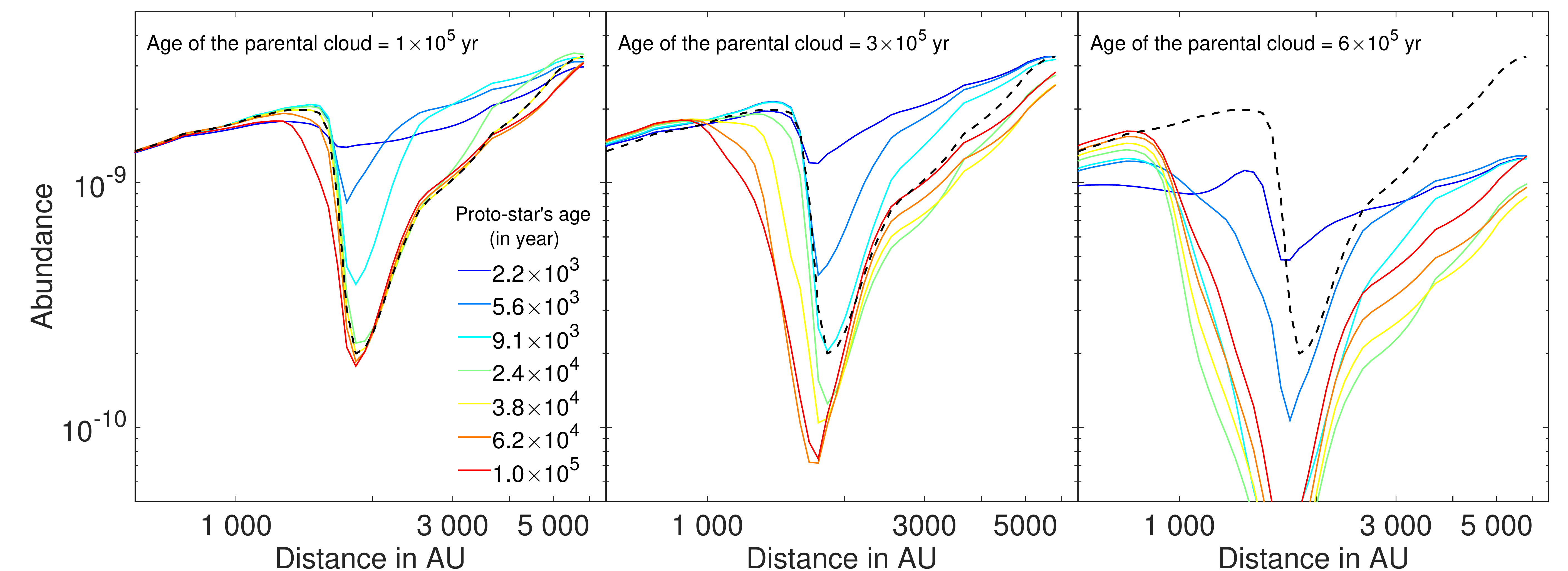}
	\includegraphics[width=0.95\hsize, clip=true, trim= 30 72 0 12]{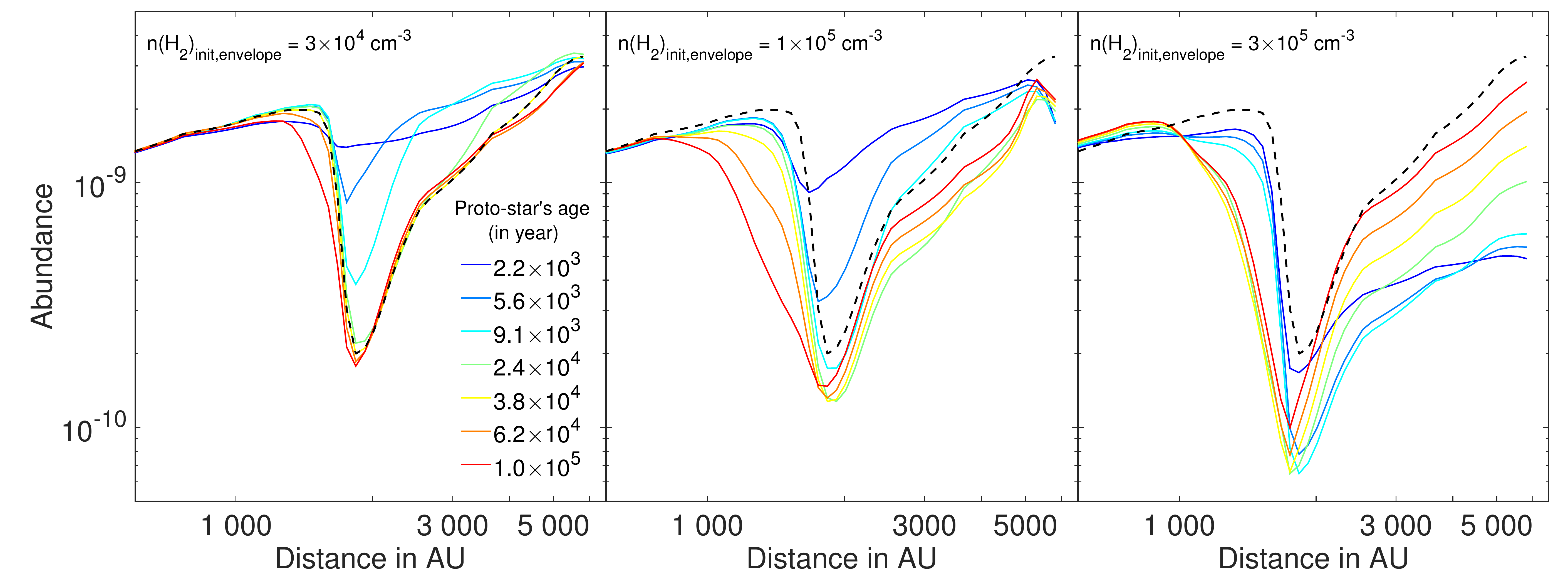}
	\includegraphics[width=0.956\hsize, clip=true, trim= 20 0 0 12]{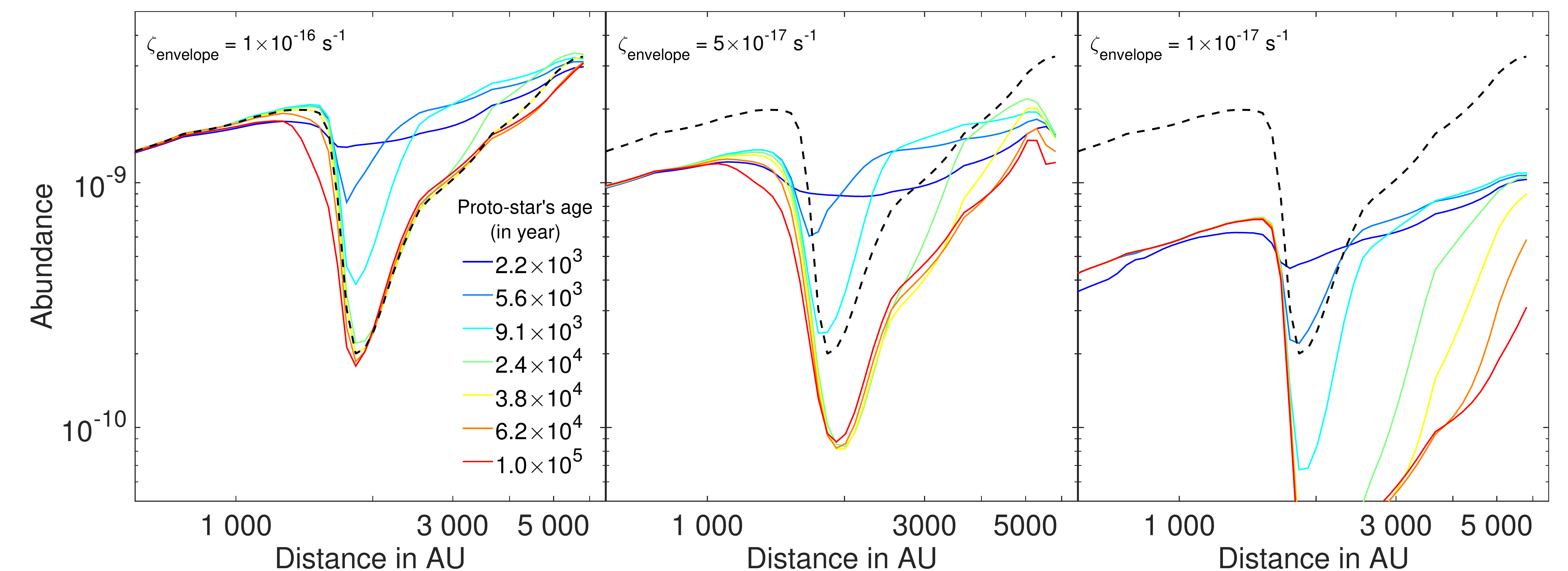}
	\caption{Variation of the abundance profile of \hco determined by \textsc{nautilus} as a function of the radius for different ages of the proto-star and different set of input chemical parameters. The best fit (age of the parental cloud = $1\times10^5$\,yr, $n({\rm H_2})=3\times10^{4}$\,cm$^{-3}$, cosmic ray ionisation rate = 1$\times10^{-16}$\,s$^{-1}$, and age of the proto-star = $3.8\times10^4$\,yr) is shown in black dashed lines. \textit{Top panels:} Variation of the age of the parental cloud (from left to right): $1\times10^5$, $3\times10^5$, and $6\times10^5$\,yr. \textit{Middle panels:} Variation of the initial H$_2$ density in the parental cloud (from left to right): $3\times10^{4}$, $1\times10^{5}$, and $3\times10^{5}$\,cm$^{-3}$. \textit{Bottom panels:} Variation of the cosmic ray ionisation rate (from left to right): $1\times10^{-16}$, $5\times10^{-17}$, and $1\times10^{-17}$\,s$^{-1}$.}
	\label{chem_hco}
\end{figure*}

\begin{figure*}
	\centering
	\includegraphics[width=0.8\hsize, clip=true, trim=0 0 0 0]{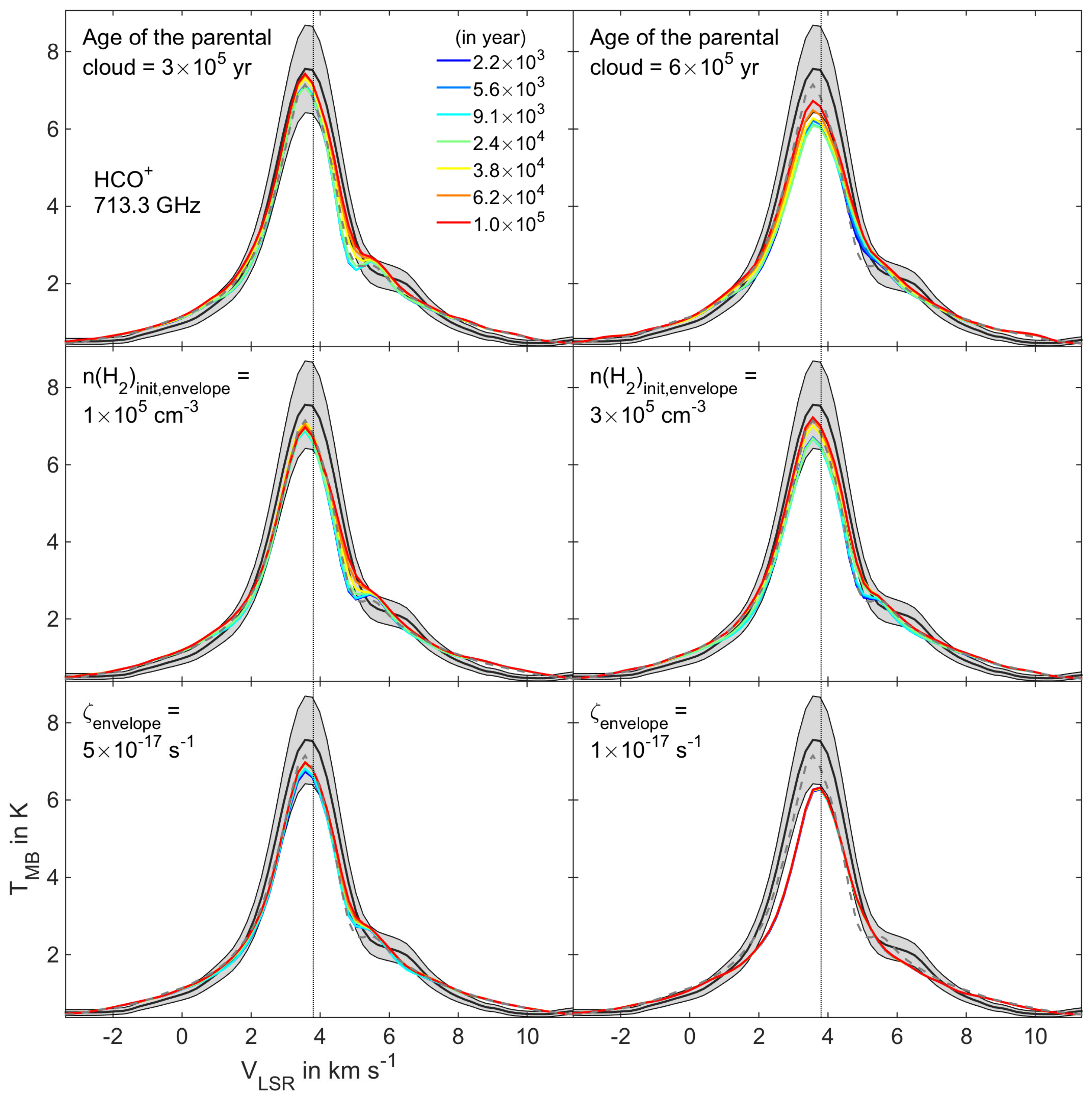}
	\caption{Line profiles of the \hco(8--7) transition for different abundance profiles determined by \textsc{nautilus} for different ages of the proto-star and different sets of input chemical parameters (shown in the middle and right panels of Fig. \ref{chem_hco}). The best fit model is shown in grey dashed lines. The grey area shows a 20\% percent error area with respect to the observation (black solid line). The vertical black dotted line shows the supposed V$_{\rm{LSR}}=3.8$\,km\,s$^{-1}$ of IRAS16293. \textit{Top panels:} Variation of the age of the parental cloud (from left to right): $3\times10^5$, and $6\times10^5$\,yr. \textit{Middle panels:} Variation of the initial H$_2$ density in the parental cloud (from left to right): $1\times10^{5}$, and $3\times10^{5}$\,cm$^{-3}$. \textit{Bottom panels:} Variation of the cosmic ray ionisation rate (from left to right): $5\times10^{-17}$, and $1\times10^{-17}$\,s$^{-1}$.}
	\label{error_envelope}
\end{figure*}

Proto-stellar envelopes are by nature dynamical objects and the time scale of collapse may change the chemical composition of the envelopes \citep[see][]{aikawa2008, wakelam2014}. The \hco emission however seems to originate from the outer part of the envelope ($\gtrsim$\,1000\,au). Indeed, at this radius, the gas-phase abundance of H$_3^+$ is enhanced due to the cosmic-ray ionisation (see Sect. \ref{chem_hco}) and it reacts with CO, increasing the abundance of HCO$^+$. In this region the physical conditions are evolving much more slowly, and, for this reason, the use of a static model to derive the \hco abundance in the envelope rather than a dynamical structure is justified here.

Five different parameters have been separately varied (age, temperature, and density of the parental cloud, cosmic ray ionisation rate, and age of the proto-star) to discriminate the impact of the chemical modelling input parameters on the radial abundance profile of \hco\!. We have then used this radial abundance profile to predict the line emission of \hco in order to compare with observations. We emphasise that the tested models also include the outflow and the foreground cloud structure. The best model (determined following the method described in Sect. \ref{methods}) gives a parental cloud evolving for $1\times10^5$\,yr with an initial gas density $n({\rm H_2})=3\times10^{4}$\,cm$^{-3}$, and a cosmic ray ionisation rate of 1$\times10^{-16}$\,s$^{-1}$ with the age of the proto-star estimated to be $3.8\times10^4$\,yr.
Fig. \ref{chem_hco} shows the effect of the variation of the chemical parameters on the radial abundance profile of \hco\! and Fig. \ref{error_envelope} the resulting line profile of the \hco(8--7) transition.\\

\textit{Age of the parental cloud.} It is well constrained to be $\leqslant3\times10^5$\,yr since for older ages the amount of \hco drops drastically (see top panels of Fig. \ref{chem_hco}) therefore the predicted \hco emission is weaker by more than 20\% compared to the observations (see top panels of Fig. \ref{error_envelope}).\\

\textit{H$_2$ density of the parental cloud.} It is poorly constrained but a higher value of the H$_2$ density leads to a lower abundance in the external part of the envelope (see middle panels of Fig. \ref{chem_hco} and Fig. \ref{error_envelope}). A smaller abundance in this region of the envelope reduces the self-absorption feature of low upper energy transitions ($J_{{\rm up}}<4$). If the density is not too high ($\leqslant1\times10^{5}$\,cm$^{-3}$), the \hco abundance varies by less than $\sim10\%$ at 2000\,au, keeping the integrated flux of the lines within our threshold of 20\%.\\

\textit{Kinetic temperature of the parental cloud.} A variation of this parameter does not change significantly the resulting \hco abundance profile (and hence the line profiles) so it was arbitrarily fixed to 10\,K, according to constraints given by previous studies \citep[e.g.][]{bottinelli2014, wakelam2014}.\\

\textit{Cosmic ray ionisation rate.} The ionisation rate is strongly constrained by the chemical modelling since it drastically affects the amount of HCO$^+$ in the envelope. A lower cosmic ray ionisation rate reduces the amount of \hco (see bottom panels of Fig. \ref{chem_hco}) produced throughout the source as well as the intensity of the lines (see bottom panels of Fig. \ref{error_envelope}). A rate larger than $\sim5\times10^{-17}$\,s$^{-1}$ is necessary, otherwise intensities of the modelled lines are below the 20\% limit. The rate is higher than the standard value of 1.3$\times10^{-17}$\,s$^{-1}$ found in the solar neighbourhood but the $\rho$\,Ophiuchus cloud complex is known for its high cosmic ray rate \citep{hunter1994}. This value is also consistent with previous studies reported for IRAS16293 (e.g. \citealp{doty2004} and \citealp{bottinelli2014}).\\

\textit{Age of the proto-star.} At older proto-stellar age, a drop in the \hco radial abundance arises at $\sim$\,2000\,au (caused by CO being depleted onto grain surfaces), leading to weaker self-absorption of high-J lines. Hence, the line profiles are slightly more compatible with the observations for higher proto-star ages than lower ones. However, from Fig. \ref{error_envelope}, it is difficult to give any stringent constraints on the age of the source. Taking into account the fluxes of all \hco lines, we derived a lower limit of $\sim2\times10^4$\,yr. This age is compatible with recent studies performed toward this source \citep{majumdar2016, quenard2018}.\\

We also would like to emphasise that the contribution of the envelope to the emission of \hco clearly does not dominate (see Sect. \ref{param_outflows}), therefore it is difficult to constrain the chemical input parameters. Nonetheless, some parameters such as the cosmic ray ionisation rate or the age of the parental cloud have an important impact on the abundance profile of \hco\!, and hence on the resulting line profiles, and it is possible to give good constraints on their value. For the density and temperature of the parental cloud as well as the age of the proto-star, no stringent conclusions can be drawn because the effects of these parameters on the line profiles are poorly constrained by the observations.

Finally, the structure of IRAS16293, as revealed by numerous interferometric observations, is in reality much more complicated because it is not homogeneously distributed and peak emissions of some species may occur in a specific region of the source and it can be hardly modelled, even in 3D. This effect has already been observed by e.g. \citet{jorgensen2011} and it can play an important role in the emission seen with single-dish telescopes. It can explain the difference we get between the predicted model and the observation for optically thin molecules such as \hcoo or \dcco\!. Such effects can also explain the excess in emission seen at red velocities for the \hco(1--0) transition that we struggle to perfectly reproduce (see following Section).

%________________________________________________________________

\subsection{Physical parameters of the foreground cloud}\label{phy_chem_PDR}

\begin{figure*}
	\centering
	\includegraphics[width=\hsize, clip=true, trim= 0 0 0 0]{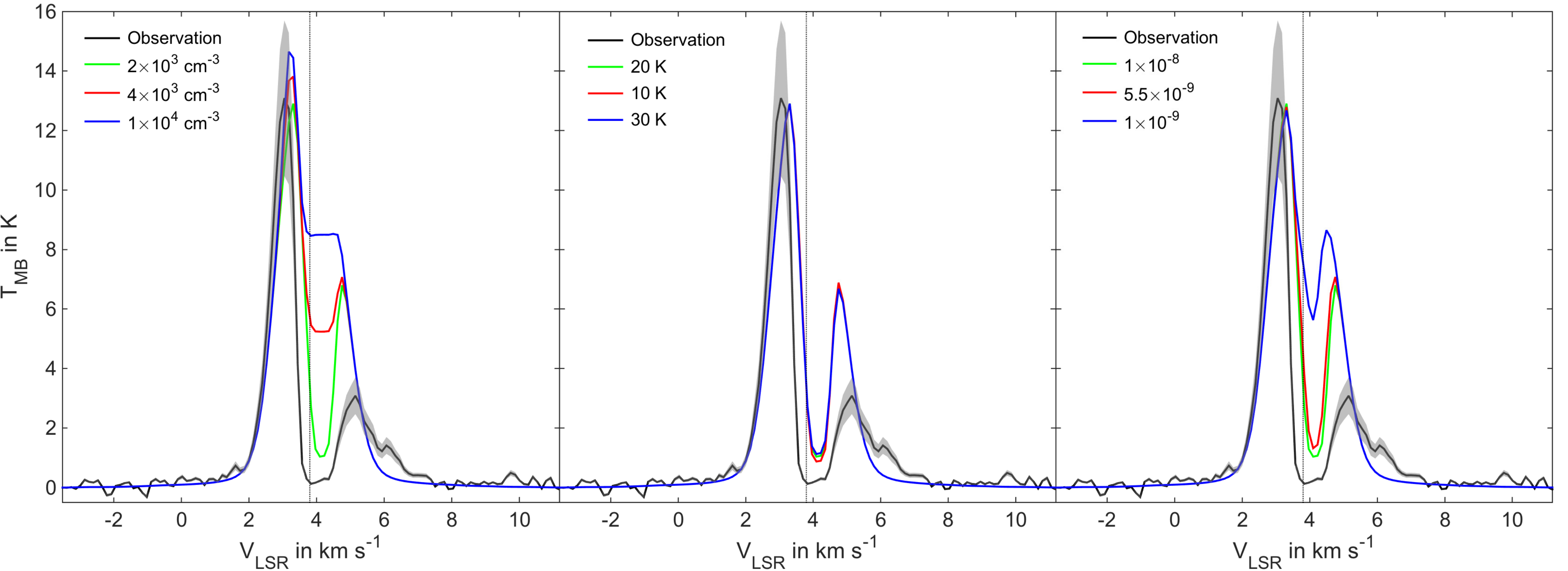}
	\caption{Line profiles of the \hco(1--0) transition for different input parameters of the foreground cloud structure. The grey area shows a 20\% percent error area with respect to the observation (black solid line). The vertical black dotted line shows the supposed V$_{\rm{LSR}}=3.8$\,km\,s$^{-1}$ of IRAS16293. The reference model (in green) is the best fit with $n({\rm H_2})_{{\rm foreground}}=2\times10^3$\,cm$^{-3}$, $T_{{\rm kin, foreground}}=20$\,K, and X(HCO$^+$)$_{{\rm foreground}}=1\times10^{-8}$. \textit{Left panel:} Variation of the H$_2$ density of the foreground cloud. \textit{Middle panel:} Variation of the kinetic temperature of the foreground cloud. \textit{Right panel:} Variation of the \hco abundance of the parental cloud.}
	\label{pdr_param}
\end{figure*}

\begin{figure*}
	\centering
	\includegraphics[width=\hsize, clip=true, trim= 0 0 0 0]{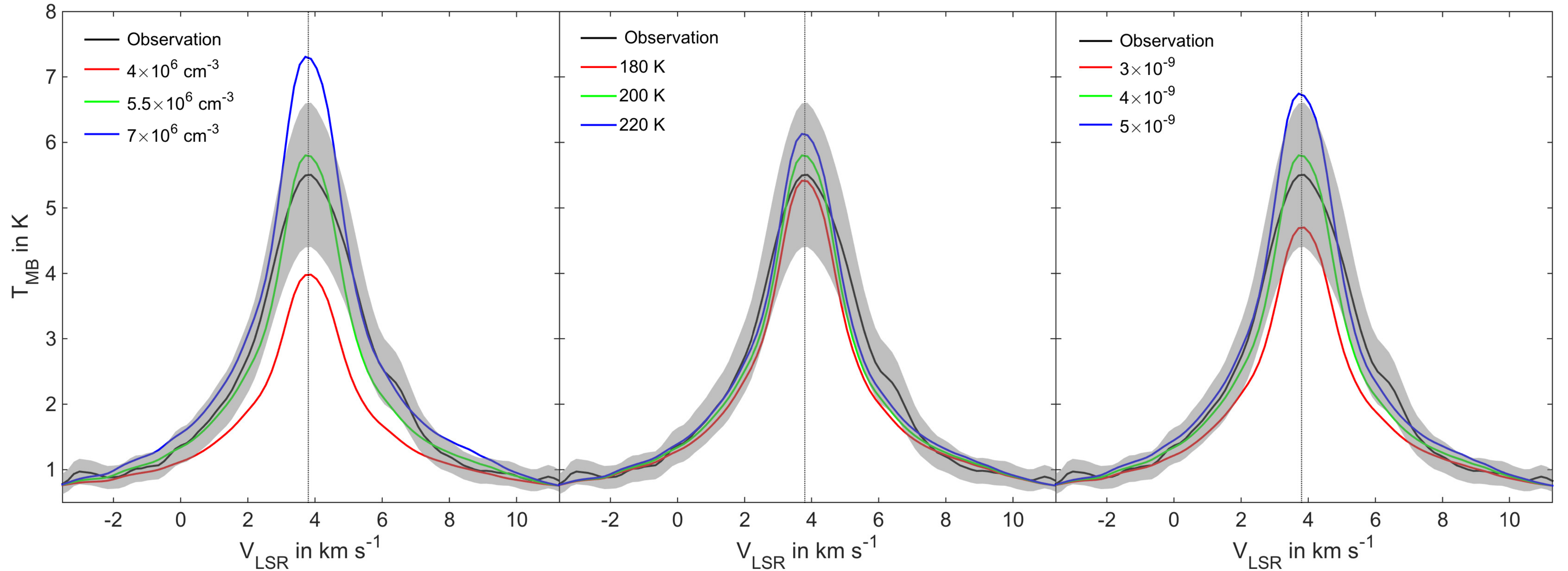}
	\caption{Line profiles of the \hco(10--9) transition for different input parameters of the foreground cloud structure. The grey area shows a 20\% percent error area with respect to the observation (black solid line). The vertical black dotted line shows the supposed V$_{\rm{LSR}}=3.8$\,km\,s$^{-1}$ of IRAS16293. The reference model (in green) is the best fit with $n({\rm H_2})_{{\rm outflow}}=5.5\times10^6$\,cm$^{-3}$, $T_{{\rm kin, outflow}}=200$\,K, and X(HCO$^+$)$_{{\rm outflow}}=4\times10^{-9}$. \textit{Left panel:} Variation of the H$_2$ density of the outflow. \textit{Middle panel:} Variation of the kinetic temperature of the outflow. \textit{Right panel:} Variation of the \hco abundance of the outflow.}
	\label{outflow_param}
\end{figure*}

The best model parameters of the foreground cloud are $n({\rm H_2})_{{\rm foreground}}=2\times10^3$\,cm$^{-3}$, $T_{{\rm kin, foreground}}=20$\,K, and X(HCO$^+$)$_{{\rm foreground}}=1\times10^{-8}$. Using Eq.\,(\ref{eq_av}), we derive $A_V\simeq1.2$ for a supposed foreground cloud depth of $3\times10^4$\,au. All three parameters have been varied at the same time to determine the best fit.\\

Fig. \ref{pdr_param} presents the emission of the \hco(1--0) transition for different models varying the foreground cloud physical parameters. One can note that the emission of this transition is clearly sensitive to the foreground cloud density and abundance, despite the short range of values shown in this figure. The reference model (in green) is the foreground cloud best fit parameters (see Table \ref{table_var}) and for each panel we vary one of the parameters only and we fix the other two to the best fit value to ease the visualisation of the impact of the parameter on the resulting line emission.\\

\textit{H$_{\textit{2}}$ density of the foreground cloud.} We have tried several densities ranging from $\sim1\times10^3$ to $\sim1\times10^5$\,cm$^{-3}$ as suggested by \citet{coutens2012} for this region combined with several kinetic temperature and molecular abundances. The density we derive ($n({\rm H_2})_{\rm{foreground}}=2\times10^3$\,cm$^{-3}$) is lower than the one used by \citet{bottinelli2014} and \citet{wakelam2014} ($n({\rm H_2})_{{\rm foreground}}=1\times10^4$\,cm$^{-3}$) but these two studies only tested two different densities ($n({\rm H_2})_{\rm{foreground}}=1\times10^4$ and $1\times10^5$\,cm$^{-3}$). By varying the abundance and density at the same time, we have found a degeneracy between the two parameters. It means that a viable solution can be found with a higher density if the abundance is lower (and vice versa). In any case, we have derived that for a density higher than $n({\rm H_2})_{{\rm foreground}}=1\times10^4$\,cm$^{-3}$ and for any abundance value, the \hco(1--0) transition is not self-absorbed enough (absorption depth higher by a factor of 20\% compared to the observations). This strongly constrains the density of the foreground cloud and its visual extinction.\\

\textit{Kinetic temperature of the foreground cloud.} Using the constraint obtained by \citet{bottinelli2014} and \citet{wakelam2014}, who determined the kinetic temperature to be $\leq30$\,K (see Sect. \ref{pdr_chem}), we tested values in the range $10-30$\,K. The line profiles do not change significantly (less than 5\% compared to one another) within this range, so we arbitrarily set the best model value to 20\,K.\\

\textit{Abundance of the foreground cloud.} The \hco abundance of $1\times10^{-8}$ we obtain is consistent with the results predicted by \citet{hartquist1998} and \citet{savage2004} at low $A_V$ for diffuse or translucent clouds. It is $\sim$2 times higher than the observed value of $[3-6]\times10^{-9}$ derived by \citet{lucas1996} in different diffuse clouds, which is a good agreement. Unfortunately, as mentioned above, there is a degeneracy between the H$_2$ density and the \hco abundance, limiting the constraints we can give to the \hco abundance ($1\times10^{-9}-1\times10^{-7}$).\\

Finally, we emphasise that none of our models satisfactorily fits the observations very well near the $+5$\,km\,s$^{-1}$ emission and the $+4$\,km\,s$^{-1}$ dip. Indeed, the \hco(1--0) transition has a very low upper energy level (4\,K) and it can be easily excited in cold conditions. As shown above, the foreground cloud physical conditions are crucial to explain the emission of this line but, in ordered to reduce the number of free parameters, we have simply considered constant physical conditions in this cloud. Further investigations (beyond the scope of this work) of the temperature and density profiles in this environment may help to recover the observed line shape.

%________________________________________________________________

\subsection{Physical parameters of the outflow}\label{param_outflows}

The best model parameters of the outflow are $n({\rm H_2})_{{\rm outflow}}=5.5\times10^6$\,cm$^{-3}$, $T_{{\rm kin, outflow}}=200$\,K, and X(HCO$^+$)$_{{\rm outflow}}=4\times10^{-9}$. All three parameters have been varied at the same time to determine the best fit.\\

Fig. \ref{outflow_param} shows the emission of the \hco(10--9) transition for different models varying the outflow physical parameters. As for the foreground cloud, the emission is clearly sensitive to the outflow density and abundance, even for the short range of parameters presented here. Plots shown in Fig. \ref{outflow_param} are presented the same way as for the foreground cloud (see \S2 of Sect. \ref{phy_chem_PDR}).\\

\textit{H$_{\textit{2}}$ density of the outflow.} A similar degeneracy as the one found for the foreground cloud has been obtained for the outflow, between density and abundance. This effect limits the constraints we can give on these two parameters. Nonetheless, to reduce the degeneracy, we have considered previous constraints on the density derived by \citet{rao2009} and \citet{girart2014} for the same object ($n({\rm H_2})\sim1\times10^7$\,cm$^{-3}$). Doing so and by varying the abundance and density at the same time, we determined a local minimum around the value of $5.5\times10^6$\,cm$^{-3}$. A little variation of the density by a factor of $\sim$30\% around this minimum leads to a difference $\gtrsim20\%$ in the predicted line fluxes, compared to the observations (see left panel of Fig. \ref{outflow_param}). Since this minimum (i) can be constrained by our observations and (ii) is consistent with previous observations, we decided to only consider it in our results.\\

\textit{Kinetic temperature of the outflow.} A lower kinetic temperature (<\,180\,K) decreases by $\sim$6\% the emission of high upper energy level lines since the gas is not hot enough to excite these transitions. On the contrary, a higher kinetic temperature (>\,220\,K) will increase the line emission by a similar factor. This effect is even more visible for $J_{\rm{up}}$ higher than the \hco(10--9) transition shown in Fig. \ref{outflow_param}. For instance, for the \hco(13--12) transition, this factor can reach a value of up to $\sim50\%$, showing how sensitive these lines are to the kinetic temperature.\\

\begin{figure}
	\centering
	\includegraphics[width=\hsize, clip=true, trim= 0 0 0 0]{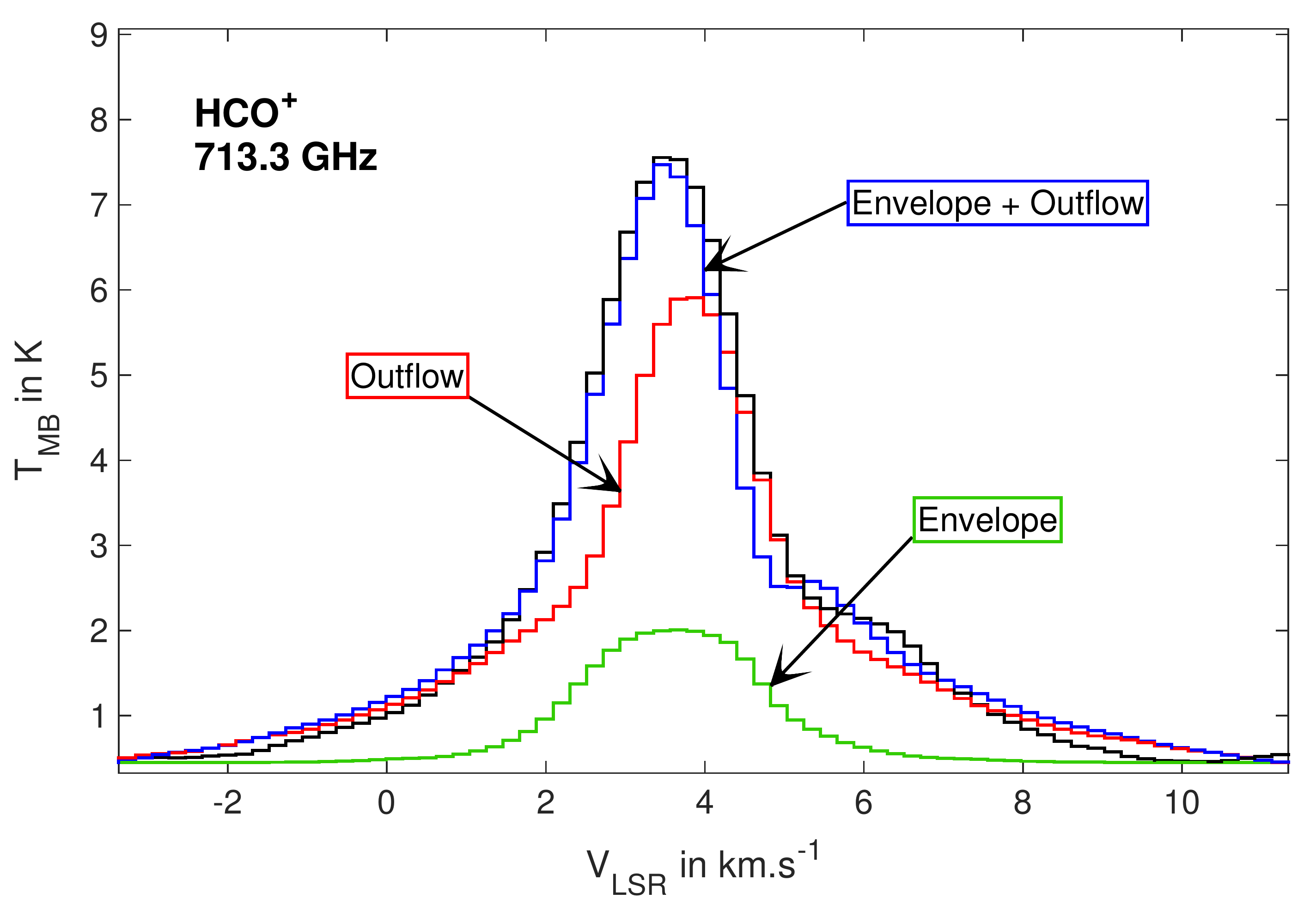}
	\caption{Line profile of the \hco(8--7) transition given by the radiative transfer modelling considering the best fit parameters of the envelope (in green) and the outflow (in red) structure and for the envelope + outflow together (in blue). For the blue curve, the radiative transfer modelling is performed with both the outflow and envelope structure in the 3D model, thus the line profile is different from a simple summation of the red and green curves. The observed line profile is plotted in black.}
	\label{outflow_envelope_emission}
\end{figure}

\textit{Abundance of the outflow.} The \hco abundance is hardly constrained due to the degeneracy with the density. However, considering the local minimum of the density, we infer a best fit \hco abundance of $\sim4\times10^{-9}$. This value is consistent with the expected enhancement of \hco in outflows described in Sect. \ref{outflow_model}. The outflow is the dominant structure in inner regions (<$8\arcsec$) for high$-J$ transitions and it largely contributes to the total emission since the envelope temperature is too low ($\sim$50\,K) to reproduce alone the emission of these lines. Moreover, when the envelope temperature is high enough ($\sim$150\,K at $\sim$40\,au), \hco is destroyed by recombination with H$_2$O and the \hco abundance becomes so low ($\sim1\times10^{-14}$) that it cannot participate in the emission of these high$-J$ transitions. This conclusion is highlighted in Fig. \ref{outflow_envelope_emission} where the contribution of the envelope and the outflow to the total emission of the \hco(8--7) transition is shown, considering the physical parameters of the best fit model. From this figure, it is clear that the contribution of the envelope (in green) to the total emission of \hco transitions is rather small (especially for high$-J$ transitions). The outflow contribution (in red) clearly dominates the total emission (shown in blue), proving the importance of the outflow structure to the emission of high$-J$ \hco lines. This statement cannot be as easily drawn for lower$-J$ lines where the emission is a result of the three structures.

We have compared our outflow results with the L1157 molecular outflow and especially with the B1 shock region. The abundance of \hco in this region has been determined by \citet{bachiller1997} and \citet{podio2014}. In their case, they do not derive any evident enhancement of \hco with respect to the dense core value \citep{podio2014}. This conclusion is consistent with previous observations toward this object \citep{hogerheijde1998, tafalla2010}. From the theoretical studies of \citet{rawlings2004} and \citet{rollins2014}, it is shown that after an age of a few hundred years and especially after $\sim500$\,yr the \hco abundance drastically drops in an outflow, and that no enhancement of its abundance with respect to the surrounding envelope is expected at older ages. This conclusion is in agreement with the observation of L1157-B1 for which the inferred shock age is t$_{\rm shock}\sim2000$\,yr.  However, in our case, the estimated dynamical age of the young NW-SE outflow is estimated to be $\sim400$\,yrs which is approximately the age for which we expect the abundance to start decreasing. This might be the case here, since the outflow abundance we derive [X(HCO$^+$)$\rm _{outflow}\sim4\times10^{-9}$] is roughly the same as the peak abundance of X(HCO$^+$)$\rm _{envelope}\sim2\times10^{-9}$ we derive for the envelope.

The ranges of outflow density and temperature we obtain in our study are consistent, within a factor of two, with \citet{rao2009} and \citet{girart2014}. However, fitting several \hco transitions, we provide a better estimation of the outflow physical conditions, particularly for the kinetic temperature.\\
 
 Using the outflow physical parameters derived from our modelling, we have calculated several outflow properties (momentum, energy, dynamical age, outflow mass rate, momentum rate, and mechanical luminosity, see Table \ref{out_prop}) based on the definition given in \citet{dierickx2015}. The outflow mass is directly extracted from the physical model using $M_{\rm out} = \mu\,m_{\rm H}\,n(\rm H)_{\rm outflow}\,v_{outflow}$, with $\mu = 2.35$ the mean molecular weight of the gas, $m_{\rm H}$ the mass of the hydrogen atom, $n(\rm H)_{\rm outflow} = 1.1\times10^7$\,cm$^{-3}$ the outflow H density, and v$_{\rm outflow}$ the volume of the outflow. The dynamical age derived from our model ($\sim$300\,yr) is close to the value found by \citet{girart2014} ($\sim$400\,yr) from CO(3--2) observations. However, the outflow mass, mass rate and momentum rate we derive are $\sim$2 orders of magnitude higher than their values. We infer that this difference is due to the different methods used to compute these quantities, derived from the accurate outflow volume and density directly extracted from our model compared to the estimate of \citet{girart2014} derived from CO lines brightness.
 Finally, we estimated the infall rate of the surrounding envelope assuming spherical symmetry \citep{pineda2012}: $\dot{M}_{\rm inf} = 4\pi r_{\rm in}^2\,n_{\rm in}\,\mu\,m_{\rm H}\,V_{\rm in}$, where $r_{\rm in}$ (1280\,au) is the radius at which the infall velocity is $V_{\rm in}$ (1.18\,km\,s$^{-1}$) with a density $n_{\rm in}$ ($2.83\times10^6$\,cm$^{-3}$). We compare this value to the outflow mass rate and we find $\dot{M}_{\rm out}$/$\dot{M}_{\rm inf}=0.35$. This value is consistent with the typical ejection over accretion ratio of 0.1--0.3 found in young stellar objects \citep{shu1988, richer2000, beuther2002, zhang2005}.

 \begin{table}
	{
	\caption{Outflow and envelope physical properties}\label{out_prop}
	\centering
	\begin{tabular}{cc}
		\hline\hline
		Physical property 					& 	Value\\
		\hline
		Outflow mass $M_{\rm out}$			&	$0.02\,M_{\odot}$\\
		Outflow momentum $P$				&	$0.15\,M_{\odot}$\,km\,s$^{-1}$\\
		Outflow energy $E$					&	$1.54\times10^{43}$\,erg\\
		Outflow mass rate $\dot{M}_{\rm out}$	&	$6.72\times10^{-5}\,M_{\odot}$\,yr$^{-1}$\\
		Outflow force $F_{\rm out}$			&	$5.04\times10^{-4}\,M_{\odot}$\,km\,s$^{-1}$\,yr$^{-1}$\\
		Mechanical luminosity $L_{\rm out}$		&	$0.42\,L_{\odot}$\\
		Dynamical time $t_{\rm dyn}$			&	$303$\,yr\\
		\hline
		Envelope mass $M_{\rm env}$			&	$2.04\,M_{\odot}$\\
		Infall mass rate $\dot{M}_{\rm inf}$		&	$1.92\times10^{-4}\,M_{\odot}$\,yr$^{-1}$\\
		\hline
		\multicolumn{2}{l}{\textbf{Notes.} The envelope mass has been determined from}\\
		\multicolumn{2}{l}{the H$_2$ density profile and is in agreement with the}\\
		\multicolumn{2}{l}{value given in \citet{crimier2010}.}\\
	\end{tabular}
	}
\end{table}
 
%________________________________________________________________

\subsection{Fractionation}
		
To reproduce the line emission of the \hco isotopologues, we varied the different isotopic ratios. We set a step of 10 for $^{16}$O$/^{18}$O and a step of 1 for $^{12}$C$/^{13}$C, and explored ranges of [300--800] and [30--80], i.e. bracketing typical local ISM values \citep{wilson1994, bensch2001}. The $^{12}$C$/^{13}$C and $^{16}$O$/^{18}$O best ratios we derived in this study are consistent with values found in the ISM \citep{wilson1994} and for the $\rho$\,Ophiuchus cloud \citep{bensch2001}. The following error bars are given for a 20\% difference between the modelled and observed line fluxes.\\

A constant $^{16}$O$/^{18}$O$=460\pm50$ ratio is sufficient to reproduce the \hcoo observations within a difference of 20\% on the line fluxes. Note that the \hcoo transition at 255.5\,GHz has been ignored in the error calculation due to a bad calibration of the IRAM 30\,m observations, as suggested by \citet{caux2011}. This ratio is in good agreement with the typical ratio of $560\pm25$ observed in the ISM \citep{wilson1994}. \citet{smith2015} have determined a $^{16}$O$/^{18}$O in the $\rho$\,Ophiuchus region for different Class I and Class II objects. From their result (see their Table 6), one can note that the ratio increases for more evolved objects, up to a few thousands. In the case of IRS\,63, a Class I object, they obtained $^{16}$O$/^{18}$O$=690\pm60$. Since IRAS16293 is a Class 0, we might expect a lower ratio, in agreement with our finding.

For the $^{12}$C$/^{13}$C ratio, we derive $51\pm5$, slightly lower (but still consistent within the error bars) than the value of $65\pm11$ found by \citet{bensch2001} in the core C of the $\rho$\,Ophiuchus molecular cloud, $\sim$2 degrees away from IRAS16293.\\

While there is no evidence for varying $^{12}$C$/^{13}$C and $^{16}$O$/^{18}$O ratios in proto-stellar envelopes, it is not the case for the H/D ratio, whose value is found to be different in different parts of the envelope \citep[e.g.][]{coutens2012}. This can be explained by the fact that, at low temperature ($T < 30$\,K) and low densities ($n({\rm H_2})$<10$^{5}$\,cm$^{-3}$), \dco is enhanced, increasing rapidly the fractionation of \hco \citep{dalgarno1984, roberts2000}. Because of the lack of observational constraints on the radial profile of the H/D ratio accross the envelope, we decided to use a pragmatic approach and to consider an \textit{ad hoc}, linear law.

For the external part of the envelope we fixed $\textnormal{H/D}=20$, the value derived by \citet{coutens2012} for the foreground cloud using water lines (HDO/H$_2$O$~\sim4.8\%$). To the best of our knowledge, this is the only estimated value of deuterium fractionation in this region of IRAS16293. In the inner region of the envelope, close to the hot corinos, the H/D ratio varies around $\sim$[30-100] for various complex organic molecules \citep{coutens2016, jorgensen2016} but \citet{persson2013, persson2014} derived a higher value ($\sim$1000) for water in the hot corino of IRAS16293 and other low-mass proto-stars. We have tried several inner value ranging between 30 and 1000 (with a step of 10) and we found that $\textnormal{H/D}=250\pm40$ gave the best match between model and observations.

For the envelope of IRAS16293, \citet{loinard2000} and \citet{loinard2001} found a value of 100 from column densities derived using only one low upper energy level transition of \hcco and DCO$^+$. Our values, between 20 and 250 depending on the radius, are consistent with this ratio. More recently \citet{koumpia2017} derived in NGC1333~IRAS\,4A, an other Class 0 proto-star, a similar ratio of 100 in the proto-stellar envelope, also in agreement with our values in the envelope of IRAS16293.

Finally, we want to emphasise that the outflow does not contribute much for the total emission of \hcoo and \dco\!. However, it might contribute significantly for \hcco, as shown in the maps from \citet{jorgensen2011} (see their Fig. 19). Moreover, since we are using only pointed observations (and not maps), it is hard to draw conclusions on the possible inhomogeneities and sub-structures of the envelope, possibly affecting the emission of the \hcoo and \dco lines (see also Sect. \ref{res_chem_env}).

\smallskip

\begin{figure}
	\centering
	\includegraphics[width=\hsize, clip=true, trim= 0 0 0 0]{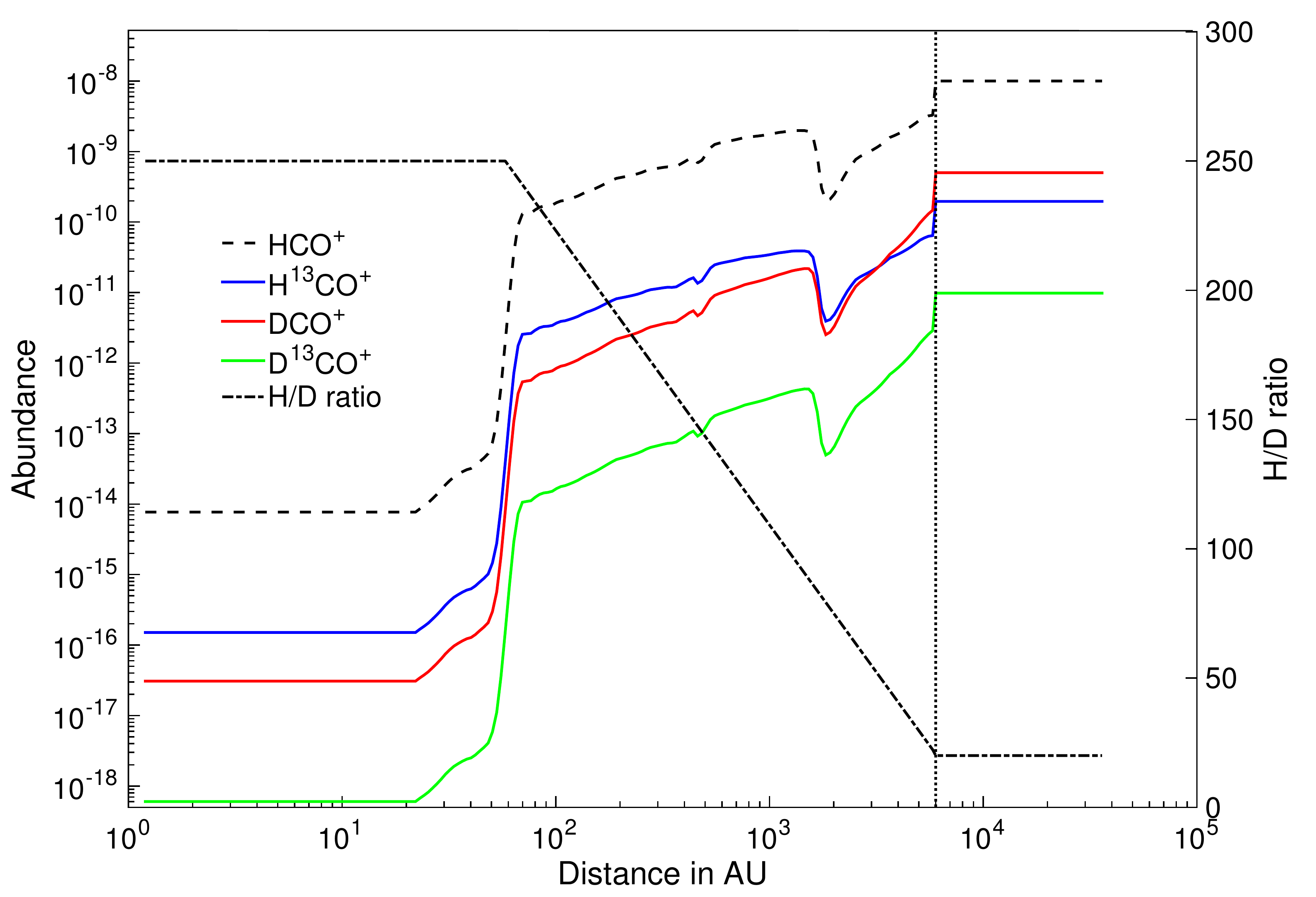}
	\caption{\textit{Left axis:} Abundance profile of \hco (black dashed line) as a function of the radius compared to the \hcco (blue), \dco (red), and \dcco (green) one. \textit{Right axis:} D/H ratio used in this study (black dash-dotted line). The vertical black dotted line shows the $\textnormal{R}=6000$\,au limit of the foreground cloud.}
	\label{plot_frac}
\end{figure}

The \hco\!, \hcco\!, \dco\!, and \dcco abundance profiles are plotted in Fig. \ref{plot_frac} alongside with the H/D ratio.

%________________________________________________________________

\subsection{Ionisation degree in the envelope}

\begin{figure}
	\centering
	\includegraphics[width=\hsize, clip=true, trim= 0 0 0 0]{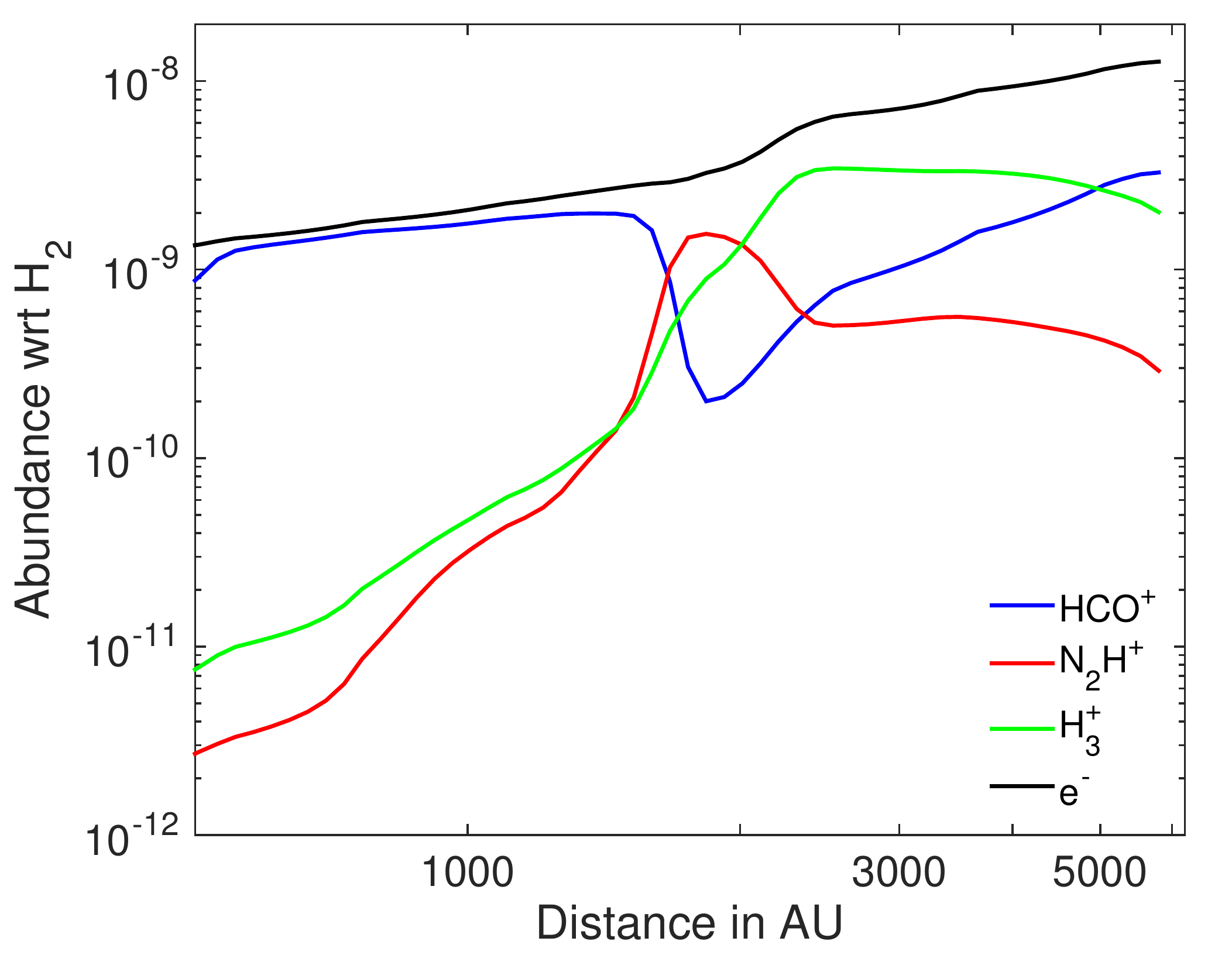}\\
	\includegraphics[width=\hsize, clip=true, trim= 0 0 0 0]{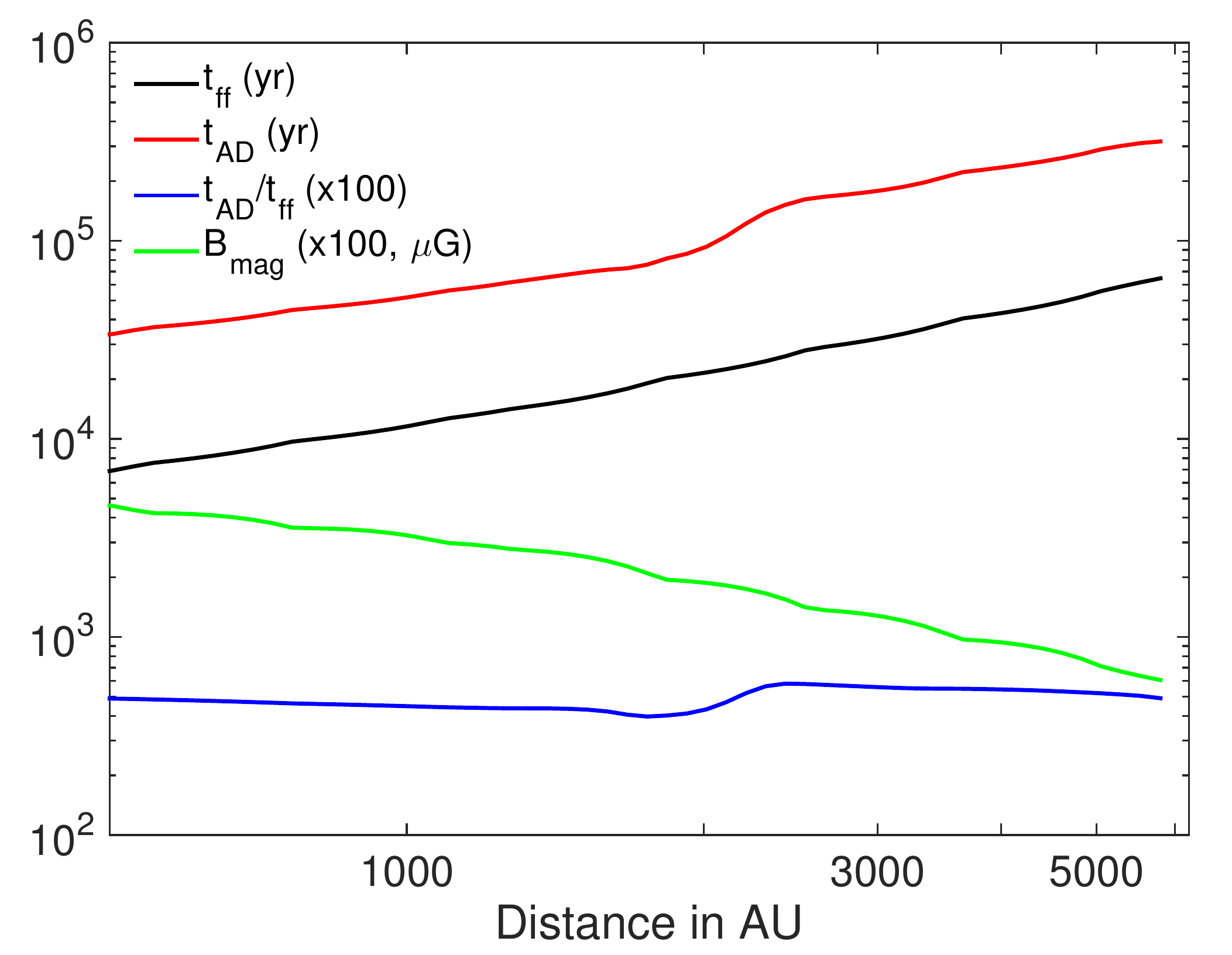}
	\caption{\textit{Top panel:} Abundance as a function of the radius for electrons (in black) and three main ions (HCO$^+$, N$_2$H$^+$ and H$_3^+$, in colours) in the envelope, extracted from the chemical modelling. \textit{Bottom panel:} Free-fall (in black) and ambipolar diffusion (in red) timescales, ratio of the two timescales (in blue, multiplied by 100) and magnetic field (in green, multiplied by 100) across the envelope.}
	\label{ionisation_plots}
\end{figure}

We have determined the ionisation degree $x(e)$ in the envelope of IRAS16293 using the results given by the chemical modelling. The ionisation degree (abundance of electron as a function of the radius) is shown in the top panel of Fig. \ref{ionisation_plots}, alongside the abundance of \hco and two other important ions in the envelope (N$_2$H$^+$ and H$_3^+$). \hco is the main ion in the envelope until $\sim$1500\,au and we have $x(e)$$\simeq$X(HCO$^+$), as expected when CO is not depleted \citep{caselli2002}. The ionisation degree we find is:
\begin{equation}
	10^{-8.9}\,\lesssim\,x(e)\,\lesssim\,10^{-7.9}.
\end{equation}
From the ionisation degree, we have been able to determine the relationship between $x(e)$ and $n(\rm H_2)$, using the envelope density profile (see Sect. \ref{chem_env}). By fitting a power-law profile, we find:
\begin{equation}\label{xenh2}
	x(e) = 3.04^{+0.60}_{-0.59}\times10^{-6}\,n(\rm H_2)^{-0.460\pm0.015}.
\end{equation}
Our value is $\sim$5 times lower than the standard value of $x(e) = 1.5\times10^{-5}\,n(\rm H_2)^{-0.5}$ determined by \citet{mckee1989} \citep[see also][]{basu1994}. This means that the ionisation balance in the envelope is not dictated by cosmic rays alone \citep{mckee1989}. \citet{caselli2002-2} also came to this conclusion in the evolved cold dense core L1544 and they inferred that the depletion of metals is the cause of the reduced ionisation degree \citep{caselli1998}, which might be also the case here. Interestingly, our $x(e) - n(\rm H_2)$ relation is close to the one obtained by \citet{caselli2002} in L1544, indicating that the physical conditions in the envelope of IRAS16293 may ressemble those of a previous pre-stellar phase, which is consistent with our choice to consider the physical conditions in the envelope static in Sect. \ref{res_chem_env}.

From the ionisation degree, we have determined the ambipolar diffusion timescale $t_{\rm AD}$ \citep{spitzer1978, shu1987}:
\begin{equation}
	t_{\rm AD}[{\rm yr}] = 2.5\times 10^{13}\,x(e).
\end{equation}
In our case, $3.4\times10^4\,{\rm yr}\,\lesssim\,t_{\rm AD}\,\lesssim\,3.2\times10^{5}$\,yr (see bottom panel of Fig. \ref{ionisation_plots}). We compared this value to the free-fall timescale $t_{f\!f}$ of the envelope, given by:
\begin{equation}
	t_{f\!f} = \sqrt{\frac{3\pi}{32\,Gn_{\rm H}\mu m_{\rm H}}}.
\end{equation}
We have $6.9\times10^3\,{\rm yr}\,\lesssim\,t_{f\!f}\,\lesssim\,6.5\times10^{4}$\,yr. Therefore, across the envelope, the $t_{\rm AD}/t_{f\!f}$ ratio remains relatively constant to $\sim$5. This ratio is a factor of two lower than the expected value for low-mass proto-star ($\sim$10, see e.g. \citealp{williams1998}). The fact that our ratio is $\sim$5 indicates that the magnetic field starts to efficiently support the envelope against gravitational collapse. In contrast, \citet{caselli2002-2} derived a value of $\sim$2 for the same ratio for the L1544 pre-stellar core, indicating that it is on the verge of dynamical collapse.

Finally, based on the definition given in \citet{nakano1972}, we derive the value of the magnetic field $B$ across the envelope:
\begin{equation}
	B[G] = 10^{-4}\times\sqrt{\frac{8\pi\mu_0\,R^2\,x(e)\,n_{\rm H}^2\,m_{\rm H}\,\langle\sigma v\rangle}{t_{\rm AD}}},
\end{equation}
with $\mu_0$ the vacuum permeability, $R$ the radial distance and $\langle\sigma v\rangle \simeq 2\times10^{-9}$cm$^{3}$\,s$^{-1}$ the average collision rate between ions and atoms \citep{osterbrock1961, basu1994}. The result is shown in the bottom panel of Fig. \ref{ionisation_plots}. The magnetic field varies between $6-46\,\mu$G, in agreement with values found in pre-stellar cores such as L1498 and L1517B \citep{kirk2006}. This shows again that the physical properties of the envelope of IRAS16293 is close to those of cold cores, consistent with the young age of the proto-star.

%________________________________________________________________

\section{Conclusions}\label{ccl}

We have used \textsc{gass} and \textsc{lime} to perform a 3D modelling of \hco and its isotopologues \hcco\!, \hcoo\!, \dco\!, and \dcco emission in the low-mass proto-star IRAS16293. We have considered an envelope, outflow, and foreground cloud structure to model the emission of these species. Thanks to the large number of detected transitions and the wide range of upper energy levels ($4-389$\,K), we have been able to study these different structures and give good constraints on the physical conditions of the outflow and foreground cloud. The contribution of the envelope to the emission of \hco clearly does not dominate, limiting the constraints we can obtain on the physical and chemical parameters. For the outflow, we have derived $T_{\rm kin}=(200\pm20)$\,K and $n({\rm H_2})=(5.5\pm1.5)\times10^6$\,cm$^{-3}$ with X(HCO$^+$)$=(4\pm1)\times10^{-9}$. The emission coming from the outflow is mostly responsible for all the J$_{\rm{up}}\geqslant8$ emission and it largely participates to other transitions. The outflow mass is 0.02\,$M_\odot$ with an outflow mass rate $\dot{M}_{\rm out} = 6.72\times10^{-5}\,M_\odot$, $\sim$3 times lower than the infall mass rate of the surrounding envelope. This value is in agreement with the typical ejection over accretion ratio of 0.1--0.3 found in young stellar objects.
We have also demonstrated that the foreground cloud causes the deep self-absorption seen for J$_{\rm{up}}\leqslant4$ lines. This is only possible if this cloud is cold ($\leqslant30$\,K) and not dense ($n({\rm H_2})\leqslant1\times10^4$\,cm$^{-3}$). We have used the chemical code \textsc{nautilus} to estimate the \hco radial abundance profile of the envelope and, combined with the outflow and the foreground cloud contributions, we have been able to reproduce correctly the observed lines. By using multiple isotopologues, we also derived several fractionation ratio: $^{12}$C/$^{13}$C$=51\pm5$, $^{16}$O$/^{18}$O$=460\pm50$, and H/D$=20-250$.
Finally, using the \hco abundance profile across the envelope, we have been able to estimate the ionisation degree to be $10^{-8.9}\,\lesssim\,x(e)\,\lesssim\,10^{-7.9}$. The derived $x(e)-n(\rm H_2)$ relation is consistent with the one found in pre-stellar cores. The ambipolar diffusion timescale is $\sim$5 times higher than the free-fall timescale, indicating that the magnetic field is playing a supporting role against the gravitational collapse of the envelope. The inferred magnetic field strength is $6-46\,\mu$G, consistent with values found in pre-stellar cores.

%________________________________________________________________

\section*{Acknowledgements}

HIFI has been designed and built by a consortium of institutes and university departments from across Europe, Canada and the United States under the leadership of SRON Netherlands Institute for Space Research, Groningen, The Netherlands and with major contributions from Germany, France and the US. Consortium members are: Canada: CSA, U.Waterloo; France: IRAP (formerly CESR), LAB, LERMA, IRAM; Germany: KOSMA, MPIfR, MPS; Ireland: NUIMaynooth; Italy: ASI, IFSI-INAF, Osservatorio Astrofisico di Arcetri-INAF; Netherlands: SRON, TUD; Poland: CAMK, CBK; Spain: Observatorio Astronómico Nacional (IGN), Centro de Astrobiología (CSIC-INTA). Sweden: Chalmers University of Technology - MC2, RSS \& GARD; Onsala Space Observatory; Swedish National Space Board, Stockholm University - Stockholm Observatory; Switzerland: ETH Zurich, FHNW; USA: Caltech, JPL, NHSC. We thank many funding agencies for financial support. DQ acknowledges the financial support received from the STFC through an Ernest Rutherford Grant and Fellowship (proposal number ST/M004139). VW thanks the French CNRS/INSU programme PCMI and the ERC Starting Grant (3DICE, grant agreement 336474) for their funding.



\bibliographystyle{mnras}
\bibliography{All_ref} % if your bibtex file is called example.bib

%%%%%%%%%%%%%%%%%%%%%%%%%%%%%%%%%%%%%%%%%%%%%%%%%%

%%%%%%%%%%%%%%%%% APPENDICES %%%%%%%%%%%%%%%%%%%%%

\appendix

\section{Line profiles}
In this Appendix we present the line profiles of \hco\!, \hcco\!, \hcoo\!, \dco\!, and \dcco\! observed transitions compared to the best fit model. Please note that the observed \hcoo transition at 255.5\,GHz is badly calibrated, as suggested by \citet{caux2011}, explaining the large difference between the observation and the model. Differences between the model and the observation line profiles for low$-J$ transitions ($\lesssim$350\,GHz) may be due to inhomogeneities in the envelope structure (see Sect. \ref{res_chem_env}).

The \hco(3--2) and (4--3) lines, observed at APEX, are clearly much brighter than the other lines, and their intensities cannot be reproduced by our model that reproduces well all other lines. We suspect an intensity calibration problem for these two lines, for which we nevertheless correctly reproduce the line profiles.

\begin{figure*}
	\centering
	\includegraphics[width=0.95\hsize, clip=true, trim= 0 0 0 0]{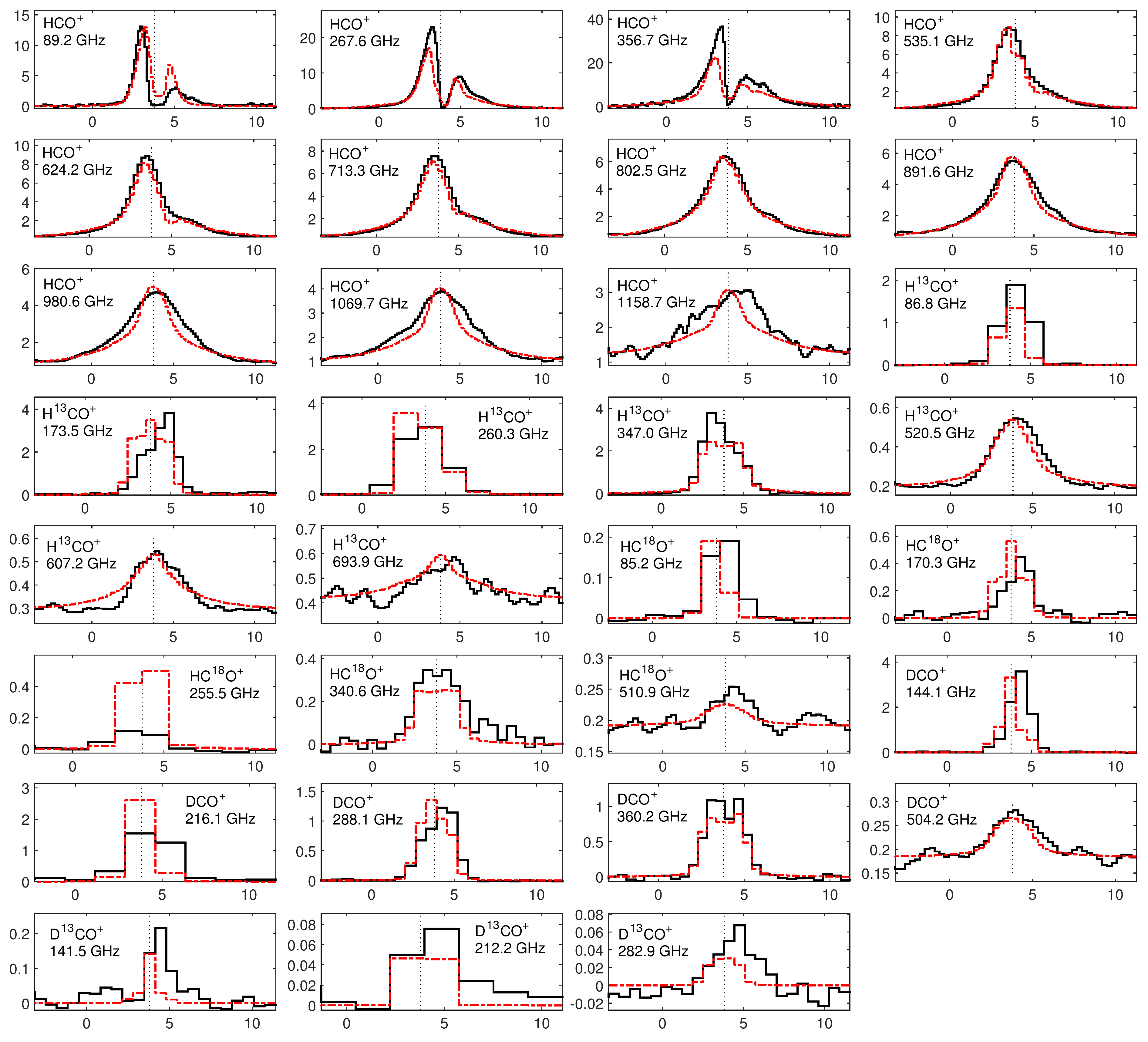}
	\caption{Main beam temperature (in K) of \hco\!, \hcco\!, \hcoo\!, \dco\!, and \dcco\! observed transitions (in black) compared to the best fit model (in red) as a function of the velocity (in km\,s$^{-1}$). The vertical black dotted line shows the V$_{\rm{LSR}}=3.8$\,km\,s$^{-1}$ of IRAS16293. Note that the observed \hcoo transition at 255.5\,GHz is badly calibrated, as suggested by \citet{caux2011}.}
	\label{best_fit}
\end{figure*}
    
%%%%%%%%%%%%%%%%%%%%%%%%%%%%%%%%%%%%%%%%%%%%%%%%%%

% Don't change these lines
\bsp	% typesetting comment
\label{lastpage}
\end{document}